\documentclass{ptephy_v1}%%%%%% to generate preprint number with ptep logo

\preprintnumber{XXXX-XXXX} %%% %%% Insert preprint number here

\usepackage{amssymb,epsf}
\usepackage{amsmath,amssymb}
\usepackage{graphicx}
\usepackage{setspace}
\usepackage{fancyhdr}
\usepackage{ifpdf}
\usepackage{graphicx}
\usepackage{color}
\usepackage{comment}
\usepackage{wrapfig}

\newcommand{\Slash}[1]{{\ooalign{\hfil$#1$\hfil\crcr\raise.167ex\hbox{/}}}}
\newcommand{\be}{\begin{equation}}
\newcommand{\ba}{\begin{eqnarray}}
\newcommand{\ea}{\end{eqnarray}}
\newcommand{\ee}{\end{equation}}

\begin{document}

\title{Holographic subregion complexity of a 1+1 dimensional $p$-wave superconductor}

%%%% To generate auto affiliation numbers please use \author{}\affil{} command

\author{Mitsutoshi F{\sc ujita}}
\affil{School of Physics and Astronomy, Sun Yat-Sen University, Guangzhou 510275, China \email{fujita@mail.sysu.edu.cn}}

%%% To include the collaborator name... Please use the command "\collaborator"
%%% For example: \collaborator{ATLAS Collaboration}

\begin{abstract}%
We analyze the holographic subregion complexity in a $3d$ black hole with the vector hair. This $3d$ black hole is dual to a  $1+1$ dimensional $p$-wave superconductor. We probe the black hole by changing the size of the interval and by fixing $q$ or $T$. We show that the universal part is finite across the superconductor phase transition and has competitive behaviors different from the finite part of entanglement entropy. The behavior of the subregion complexity depends on the gravitational coupling constant divided by the gauge coupling constant. When this ratio is less than the critical value, the subregion complexity increases as temperature becomes low. This behavior is similar to the one of the holographic $1+1$ dimensional $s$-wave superconductor arXiv:1704.00557. When the ratio is larger than the critical value, the subregion complexity has a non-monotonic behavior as a function of $q$ or $T$. We also find a discontinuous jump of the subregion complexity as a function of the size of the interval. The subregion complexity has the maximum when it wraps the almost entire spatial circle. Due to competitive behaviors between normal and condensed phases, the universal term in the condensed phase becomes even smaller than that of the normal phase by probing the black hole horizon at a large interval. It implies that the formed condensate decreases the subregion complexity like the case of the entanglement entropy. 
\end{abstract}

\subjectindex{xxxx, xxx}

\maketitle

\section{Introduction}
 Entanglement entropy is an important non-local quantity in quantum information, capturing geometric aspects of field theories (e.g. an area law~\cite{EE} and the strong subadditivity). The entanglement entropy counts the number of degrees of freedom in the quantum entangled state~\cite{Cardy,Cardy2,Review}, while it turns out to be an order parameter of the phase transition as in the Wilson loop operator in gauge theories (see quantum critical phase transitions~\cite{Vidal:2002rm}). 
Duality between strongly coupled gauge theories and the weakly coupled gravity called the gauge/gravity correspondence~\cite{Maldacena:1997re} has been a powerful tool to analyze the entanglement entropy.~\footnote{
Non-local quantities such as the Wilson loop operator were analyzed by using the minimal surface of the string worldsheet in the gauge/gravity correspondence~\cite{Maldacena:1998im}.}
The gravity dual to the entanglement entropy is given by the minimal surface called Ryu-Takayanagi surface \cite{Ryu:2006bv,Ryu:2006ef,Nishioka:2009un}. Ryu-Takayanagi surface is a useful way of analyzing the entanglement entropy of strongly coupled systems. The holographic entanglement entropy has been an order parameter of the confinement/deconfinement phase transition~\cite{Nishioka:2006gr}-\cite{Bah:2008cj} and the probe of superconductor phase transitions~\cite{Albash:2012pd}-\cite{Das:2017gjy}.

Besides, the complexity in quantum information describes the minimal number of gates of any quantum circuit to obtain a desired target state from a reference state. The holographic dual of the complexity has recently been remarked. First, the holographic complexity was conjectured by Susskind in dual black holes~\cite{Susskind:2014rva}. The holographic complexity in a black hole is given by the surface of the Einstein-Rosen bridge. The holographic complexity grows linearly in time as the length of the surface grows. The complexity of a state is in proportion to the volume of the codimension-1 maximal bulk surface $V$ in general (complexity$=$volume conjecture $C\sim V/\kappa^2 l$). However, the length scale $l$ in the complexity$=$volume conjecture is unclear for separate backgrounds. On the other hand, the complexity$=$action conjecture improves the ambiguity of the length. Following this conjecture, the Einstein-Hilbert action in the Wheeler-DeWitt patch turns out to be the holographic dual of complexity, the coefficient of which will have the physical meaning~\cite{Brown:2015bva,Brown:2015lvg}.~\footnote{Moreover, an optimization way of doing path integrals has been proposed to understand complexity in quantum field theory~\cite{Caputa:2017yrh, Bhattacharyya:2018wym}. } 

In this paper, we compute the holographic complexity of subregions. Namely, we evaluate the holographic complexity of the mixed state by tracing out states of a separate region. The holographic subregion complexity is proportional to the volume surrounded by the minimal surface (Ryu-Takayanagi surface)~\cite{Alishahiha:2015rta} as follows (see also a generalization~\cite{Alishahiha:2018lfv}):
\ba
C=\dfrac{\mbox{Volume}(\gamma_A)}{\kappa^2 R},
\ea 
where $\gamma_A$ is an area of the extremal surface and $R$ is the radius of curvatures in the background.~\footnote{More precisely, this volume is the co-dimension one maximal volume attached to the extremal surface as well as the entangling region~\cite{Carmi:2016wjl,Ben-Ami:2016qex}.} The subregion complexity leads to a discontinuous jump at the transition, which is confirmed by computing the integration of the volume form~\cite{Ben-Ami:2016qex} and using the Gauss-Bonnet theorem~\cite{Gan:2017qkz,Abt:2017pmf}. In the context of the tensor network, numerical results in the Ising model in a squared lattice reproduce an expected linear law behavior of the holographic subregion complexity.~\footnote{Numerical results do not reproduce the finite part due to the lack of rotational symmetry in a square lattice~\cite{Abt:2017pmf}. } The holographic complexity has also been computed for probing string backgrounds~\cite{Bhattacharya:2018oeq} and anisotropic black branes~\cite{HosseiniMansoori:2018gdu}.

The motivation in this paper is further to analyze the behavior of the holographic subregion complexity across a holographic superconductor phase transition. Since the holographic subregion complexity is surrounded by Ryu-Takayanagi surface, these can probe the black hole horizon by changing the size of the interval as similar to the holographic entanglement entropy.  We are interested in the finiteness of the universal term unlike the entanglement entropy divergent at a critical point of a $2d$ quantum critical phase transition in the infinite length~\cite{Vidal:2002rm,Cardy2}. In the finite length, on the other hand, finite size effects should be taken into account. In~\cite{Zangeneh:2017tub}, the holographic subregion complexity was used as the probe through the holographic 1+1$d$ $s$-wave superconductor phase transition. By improving the result of~\cite{Momeni:2016ekm}, the universal terms of the holographic complexity are numerically shown to become finite through the $s$-wave superconductor phase transition. In the holographic $s$-wave superconductor, the universal term does not behave as in the entanglement entropy through the superconductor phase transition. On the other hand, this is not the case in the $AdS$-Schwarzschild and Reissner-Nordstrom $AdS$~\cite{Roy:2017kha}, where the behavior of the holographic complexity mimics the one of the holographic entanglement entropy in some regions. Thus, it is interesting to analyze the holographic complexity in other holographic superconductor models to show the finiteness of the universal terms and to investigate the difference from the holographic entanglement entropy furthermore.

 For the computation of the subregion complexity, we focus on a specific holographic model dual to the 1+1 dimensional $p$-wave superconductor phase transition. $3d$ $SU(2)$Yang-Mills term and the Einstein Hilbert action are dual to a 1+1 dimensional $p$-wave superconductor as proven in the probe limit~\cite{Gao:2012yw,Bu:2012qr,Peng:2017kvs}. 
%Compared with the $s$-wave superconductor, parity symmetry is spontaneously broken in the $p$-wave 
%superconductor phase transition in the probe limit. A vector condensate turns out to be the order parameter of the phase transition. 
In the large $N$ limit, one can evade Coleman-Mermin-Wagner theorem in this lower dimensional system~\cite{Anninos:2010sq}: quantum fluctuations preventing formation of condensates are suppressed in the large $N$ limit. The holographic entanglement entropy is computed in a fully backreacted metric of a $1+1$ dimensional $p$-wave superconductor across the phase transition~\cite{Das:2017gjy}. The backreacted metric turns out to be a black hole with the vector hair in the condensed phase, while it turns out to be the $AdS_3$ charged black hole in the normal phase.  It is shown that the order of the $p$-wave superconductor phase transition varies depending on the strength of the coupling constant (see appendix \ref{FRE}).

In this paper, we compute the holographic subregion complexity in a fully backreacted metric of a $1+1$ dimensional $p$-wave superconductor. We make use of the divergent form of the holographic complexity analyzed in~\cite{Carmi:2016wjl,Bakhshaei:2017qud}. After analyzing the coefficient of the divergent term by varying the size of the subregion, we specify the size dependence of this coefficient. Subtracting the divergent term, we analyze the finite part of the subregion complexity.  The finite part of the subregion complexity should also depend on the strength of the coupling constant. In main section, we show that the subregion complexity as a function of $T$ or $q$ suddenly jumps near the phase transition for the large ratio of the gravitational coupling constant to the gauge coupling constant. 

In section \ref{sec2}, we review $3d$ Einstein-Hilbert and $SU(2)$ Yang-Mills action, which are dual to the $1+1$ dimensional $p$-wave superconductor. To analyze the holographic subregion complexity, we compute the backreaction of the Yang-Mills term into the metric. In section \ref{sec3}, we compute both the holographic entanglement entropy and the holographic subregion complexity in the holographic $1+1$ dimensional $p$-wave superconductor phase transition. We compute the holographic subregion complexity by fixing $q$ or $T$ (or both quantities). In section \ref{REE}, we analyze the renormalized entanglement entropy as a universal term of the entanglement entropy. We compare it with the finite term of both the holographic entanglement entropy and the subregion complexity.

\section{Backreactions of the Yang-Mills term}\label{sec2}
The $SU(2)$ Yang-Mills theory for the $AdS_3$ black hole has been a holographic model of the $p$-wave superconductor~\cite{Gao:2012yw,Bu:2012qr}. In this section, we review the holographic $p$-wave superconductor in $3d$ Einstein-Hilbert action with $SU(2)$ Yang-Mills term. We consider the action of the Einstein-Hilbert and the $SU(2)$ Yang-Mills term as \ba\label{ACT11}
I_{G}=\dfrac{1}{2\kappa^2}\int d^3x\sqrt{-g}\Big(R+\dfrac{2}{L^2}\Big)-\dfrac{1}{2g_{YM}^2}\int d^3x\sqrt{-g}\mbox{tr}(F_{\mu\nu}F^{\mu\nu}), 
\ea
where the field strength is defined as $F_{\mu\nu}=\partial_{\mu}A_{\nu}-\partial_{\nu}A_{\mu}-i[A_{\mu},A_{\nu}]$. 
 {Note that this normalization of the Yang-Mills term is convenient when it is compared with the one of the Maxwell theory. That is, using $\mbox{tr}(T^aT^b)=\delta^{ab}/2$, the kinetic term is as in $F_{\mu\nu}^aF^a_{\mu\nu}/4g_{YM}^2$.} 

By performing the coordinate transformation, a general ansatz for the metric is given by 
\ba\label{MET1}
ds^2=\dfrac{L^2}{z^2}\Big(-f(z)dt^2+dy^2+\dfrac{dz^2}{h(z)f(z)}\Big),
\ea
where $y$ is compactified with the periodicity $y\sim y+2\pi L$. The function $f(z)$ is the blackening factor which gives the position of the black hole horizon at $z=z_h$. 
The ansatz for the background non-Abelian gauge field becomes in the radial gauge (see also~\cite{Gubser:2008wv,Ammon:2009xh}) 
\ba
A=\dfrac{1}{2}(\phi(z)\sigma^3dt+w(z)\sigma^1dy),\quad A_z^b=0,
\ea
where $\sigma^a$ ($a=1,2,3$) are Pauli matrices. 

The Einstein equations derived from eq. \eqref{ACT11} turn out to be following three equations:
\ba\label{EIN245}
&\dfrac{f \left(z h f'+z f h'-2 f h+2\right)}{ z^2}  \nonumber \\
&-\dfrac{\tilde{\kappa}^2 z^2 (\phi^2 w^2 + f h (\phi^{\prime 2} + f w^{\prime 2}))}{L^2}=0,  \\
&\dfrac{2 z^2 h f''+z^2 f' h'-4 z h f'-2 z f h'+4 f h-4}{2 z^2 } \nonumber \\
&+ \dfrac{\tilde{\kappa}^2 z^2 \left(\phi^2 w^2-f h \left(f w^{\prime 2}+\phi^{\prime 2}\right)\right)}{L^2 f}=0, \nonumber \\
&-\dfrac{z h f'-2 fh+2}{ z^2 f h}- \dfrac{\tilde{\kappa}^2 z^2 \left(f h \left(f w^{\prime 2}-\phi^{\prime 2}\right)+\phi^2 w^2\right)}{L^2 f^2 h}=0, \nonumber
\ea
where the last equation is the $z$-component corresponding to the constraint equation. Here, we have introduced a parameter $\tilde{\kappa}=\kappa/g_{YM}$, which has dimension $-1$.
In addition, EOM in terms of Yang-Mills fields are written as
\ba\label{EOM37}
&-\sqrt{h}f(z\sqrt{h} \phi')'+zw^2 \phi =0, \\
&\sqrt{h}f(z \sqrt{h}fw')'+z \phi^2 w=0.  \nonumber 
\ea
Due to the dependence of these EOM only on the dimensionless combination $\tilde{\kappa}/L$, $L$ is set to be 1 in remaining section.

\subsection{The normal phase}
We then solve the Einstein equation of motion derived from the action eq. \eqref{ACT11}. In the normal phase, the $y$ component of the gauge field is zero, while non-zero $A_t^3=\phi$ produces the charge density and breaks $SU(2)$ gauge symmetry into $U(1)_3$. The energy momentum tensor turns out to be those without non-linear terms. We then know the charged $AdS_3$ black hole  solution~\cite{Cadoni:2009bn,Jensen:2010em} with the unit $AdS$ radius as the solution to the Einstein equation of motion as follows: 
\ba\label{btz230}
ds^2_{normal}=\dfrac{1}{z^2}\Big(-f(z)dt^2+dy^2+\dfrac{dz^2}{f(z)}\Big), \quad \phi(z) = q\log \Big(\dfrac{z}{z_0}\Big),
\ea
where $f(z)=1-(z/z_0)^2+\tilde{\kappa}^2 q^2 z^2\log (z/z_0)$ and the black hole horizon is located at $z=z_0$. Here, $\phi(z)$ is required to be regular at the position of the horizon $z=z_0$. The squared horizon position is inversely proportional to  regularized mass $M_0=(L/z_0)^2$ which satisfies the BPS-like bound $M_0\ge \tilde{\kappa}^2 q^2/2$~\cite{Cadoni:2009bn}. The BPS-like bound is saturated at the zero temperature.  

Due to the non-normalizable log term, the gauge field obeys the alternative boundary condition, for which the charge density $q$ is considered as the source. The chemical potential turns out to be $\mu =-q\log (z_0)$.

\subsection{The condensed phase}
In this section, we consider the condensed phase, where both $\phi(z)$ and $w(z)$ are non-zero. The charged $AdS_3$ black hole is unstable when $q$ is large. The black hole acquires the vector hair to go to the stable configuration called the condensed phase. A vector operator dual to $w(z)$ condenses in the condensed phase, while it breaks parity symmetry as well as remaining $U(1)_3$ spontaneously. The critical point for the $p$-wave superconductor phase transition is determined from the scaling analysis. In the probe limit, critical charge density is $q_c=21.7T_H$~\cite{Gao:2012yw}. We analyze the order of the phase transition between the normal phase and the condensed phase by varying the coupling constant $\tilde{\kappa}$.

At the $AdS$ boundary $z\to 0$, fields are expanded as
\ba\label{BOU347}
&\phi(z)\sim q \log (z)+\mu_0 \\
&w(z)\sim W_0+v_w\log (z), \nonumber \\
&f(z)\sim \dfrac{1}{n_0}\Big(1- \dfrac{z^2}{z_0^2}\Big)+\tilde{\kappa}^2 q^2 z^2 \log (z), \nonumber\\
&h(z)\sim n_0, \nonumber
\ea
where $\mu_0$ is the chemical potential, the parameter $W_0$ is the VEV of a vector order parameter, and $v_w$ is the source conjugate to $W_0$. $z_0$ and $n_0$ are constant parameters. 

The black hole horizon is expected at $z=z_h$ and satisfying $f(z_h)=0$. The regular boundary condition $\phi (z_h)=0$ is  imposed at the black hole horizon. The analytic expansion near the black hole horizon $z=z_h$ is given by 
\ba\label{EXP18}
&\phi(z)=\alpha_1(z_h-z)+\dots, \\
&w(z)={\beta_1}+\beta_2(z_h-z)\dots \nonumber \\
&f(z)=\delta_1(z-{z_h}) +\dots \nonumber \\
&h(z)=\gamma_1+\gamma_2(z_h-z)+\dots, \nonumber   
\ea
where $(\alpha_1,\ \beta_1,\ \beta_2, \ , \delta_1,\ \gamma_1,\ \gamma_2)$ are constants. 
The Hawking temperature of this black hole solution turns out to be
\ba
T_{H}=\dfrac{1}{4\pi}{|f'(z_h)|}\sqrt{h(z_h)}=\dfrac{{|\delta_2|}\sqrt{\gamma_1}}{4\pi}.
\ea
Substituting the expansion eq. \eqref{EXP18} into the EOM eq. \eqref{EIN245} and eq. \eqref{EOM37}, we obtain 4 independent parameters $(\alpha_1,\beta_1,\gamma_1,z_h)$. Other parameters are fixed by these 4 parameters as well as $\tilde{\kappa}$.   

We then solve the EOM starting from the black hole horizon. The constraint equation of eq. \eqref{EIN245} is solved at the black hole horizon. We numerically solve first two of equations \eqref{EIN245} and equations \eqref{EOM37}, specifying regularity conditions at the horizon $z=z_h$. 

At the $AdS$ boundary, we specify the boundary conditions $W_0,\ v_w=0,\ n_0=1$. The vanishing source $v_w$ shows the superconductor boundary condition, which describes the spontaneous symmetry breaking of residual $U(1)$ symmetry generated by $A_{\mu}^3$. The superconductor boundary condition is similar to the one imposed on the charged scalar~\cite{Hartnoll:2008vx,Hartnoll:2008kx} in the holographic $s$-wave superconductor.

\begin{figure}[!h]
\centering\includegraphics[width=2.5in]{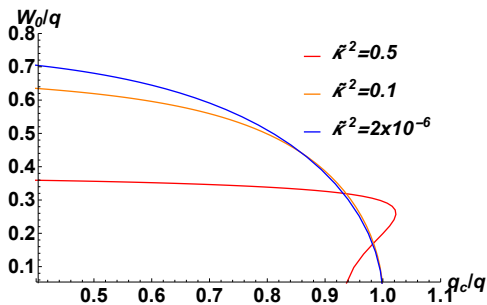}
%%%call your figure name in the place "figurename.eps"
\caption{ $W_0$ normalized by $q$ is plotted as the function of $q_c/q$ at fixed temperature $T_H=0.15$. $W_0$ increases from zero at $q=q_c$ when $\tilde{\kappa}^2<0.31$. In contrast, $W_0$ jumps to be non-zero at $q=q_c$ when $\tilde{\kappa}^2>0.31$. }
\label{fig:cond}
\end{figure}

Note that there are scaling symmetry in the EOM as follows:
\ba\label{TY350}
&(t,y,z)\to \Lambda_0^{-1} (t,y,z), \quad \phi\to \Lambda_0 \phi,\quad w\to \Lambda_0 w, \\
&f\to \Lambda_0^2 f,\quad h\to \Lambda_0^{-2} h ,\quad \phi\to \Lambda_0 \phi.
\ea
One can use first symmetry to  fix $z_h=1$. Second symmetry can be used to fix the parameter $n_0=1$, which yields the standard asymptotic $AdS_3$ metric. 

The behavior of $W_0/q$ is plotted as the function of $q_c/q$ in Fig. \ref{fig:cond} at fixed temperature $T_H=0.15$. In the figure, critical charge density $q_c$ is determined from the thermodynamic stability between normal and condensed phases. See appendix \ref{FRE}. The critical value $q_c$ depends on the coupling constant $\tilde{\kappa}$: $q_c=189T_H, \ 33.5T_H,\ 21.7T_H$ for $\tilde{\kappa}^2=0.5,\ 0.1,\ 2\times 10^{-6}$, respectively.~\footnote{When  $\tilde{\kappa}^2>0.31$, the log behavior of fields of the gravity dual would affect the scaling behavior of $q_c$.}
For all coupling constants $\tilde{\kappa}$, $W_0$ is zero at small charge density. 
 When $\tilde{\kappa}^2<0.31$, $W_0$ suddenly increases from zero at the critical density $q=q_c$. The condensate $W_0$ has the scaling behavior $\sim 1.18\sqrt{1-q_c/q}$.  This implies the second order phase transition. When $\tilde{\kappa}^2>0.31$, $W_0$ jumps to be non-zero at the critical density $q=q_c$, where $W_0$ does not follow a scaling behavior.

\section{Holographic complexity of the subregion}\label{sec3}
In this section, we compute the holographic complexity of the subregion in the holographic $d=1+1$ $p$-wave superconductor phase transition. We analyze time independent subregion complexity via holography~\cite{Alishahiha:2015rta}. We start with the metric of the 3-dimensional black hole eq. \eqref{MET1}. 
 
Recall that the holographic entanglement entropy is proportional to the area of a minimal surface $\gamma_A$. The 1-dimensional strip subregion with the size $l$ is considered. Using the metric eq. \eqref{MET1}, the embedding scalar of the surface $\gamma_A$ satisfies the EOM
\ba\label{ZPR21}
z'=\sqrt{h(z)f(z)(\dfrac{z_*^2}{z^2}-1)},
\ea
where $z=z_*$ is the turning point for the surface. Integrating the EOM, the embedding scalar turns out to be
\ba\label{XZ32}
x(z)=\int^{z_*}_{z} dz \dfrac{1}{\sqrt{h(z)f(z)\Big(\frac{z_*^2}{z^2}-1\Big)}},
\ea
$x(z)$ satisfies $x(\epsilon) =l/2$ as well as $x(z_*)=0$. Due to symmetry of the curve at the turning point $z=z_*$, the factor of 1/2 appears in front of $l$. The minimal surface ends on the particular end points. The holographic entanglement entropy is the minimal surface divided by the gravitational constant
\ba
S^{EE}=\dfrac{2\pi}{\kappa^2}(\gamma_A)=\dfrac{4\pi}{\kappa^2}\int^{z_*}_{\epsilon}dz\dfrac{1}{z\sqrt{f(z)h(z)(\frac{z^2}{z_*^2}-1)}}.
\ea
The divergent part of $S^{EE}$ is of the form $S^{EE}\sim \frac{4\pi}{\kappa^2}\log (\epsilon)$. Apart from the divergence, the finite part of the entanglement entropy $S^{EE}_{fin}$ is interesting to analyze. Note that the divergent log term depends on the regularization $\log\epsilon\to \log\epsilon -\log\Lambda_0$. To have a finite term independent of the regularization, instead, one needs to have the dimensionless combination inside the logarithm $\log \epsilon/T$. The finite part is also shifted by $\log T$, which should be just a constant and not affect the analysis.

The holographic entanglement entropy $S^{EE}_{fin}$ in the normal phase was analyzed by using charged black holes with hyperbolic horizons~\cite{Belin:2013uta} and 2nd order excitations~\cite{He:2014lfa,Momeni:2015iea}. The charge $q$ dependence of the finite part $S^{EE}_{fin}$ was analyzed in~\cite{Das:2017gjy}, when $\tilde{\kappa}^2<0.31$. The finite part $S^{EE}_{fin}$ has a cusp at the intersecting critical point between normal and condensed phases. The finite part behaves non-monotonically in the regime where the amount of the entanglement in the charge sector competes with the effect of the condensate.  When $\tilde{\kappa}^2>0.31$, $S^{EE}_{fin}$ is analyzed in appendix \ref{HEEp}. While the holographic entanglement entropy turns out to be multivalued at a region of $q<q_i$, the holographic entanglement entropy behaves similarly as in the one of  $\tilde{\kappa}^2<0.31$ at large $q>q_i$.  

Unlike the charge density $q$ dependence, we do not find any critical sizes of phase transition varying $T/T_i$. Especially, the finite part always decreases with decrease of $T/T_i$ at enough low temperature. The amount of the quantum entanglement decreases due to both decrease of temperature and the formed condensate. Thus, we do not have competition of two effects, namely, the formation of the condensate and decrease of temperature. To confirm the behavior of the entanglement entropy, we alternatively perform the computation of the renormalized entanglement entropy in section~\ref{REE}. It is finite entropy independent of the cutoff. 

 By contrast, the holographic complexity of the subregion is proposed to be proportional to the volume surrounded by the minimal surface $\gamma_A$. This subregion has the size $l$. Following~\cite{Alishahiha:2015rta}, the holographic complexity is defined as
\ba\label{COM2}
C=\dfrac{\mbox{volume}(\gamma_A)}{\kappa^2}.
\ea
Substituting the metric eq. \eqref{MET1} into the formula eq. \eqref{COM2}, the holographic complexity of the subregion turns out to be
\ba\label{COM3}
C(\epsilon)=\dfrac{c}{6\pi}\int^{z_*}_{\epsilon}\int^{x(z)}_0\dfrac{dzdx}{z^2\sqrt{f(z)h(z)}}=\dfrac{c}{6\pi}\int^{z_*}_{\epsilon}\dfrac{x(z)dz}{z^2\sqrt{f(z)h(z)}},
\ea 
where the central charge is defined as $c=12\pi /\kappa^2$. 

By using scaling symmetry of the first line in eq. \eqref{TY350}, the physics parameters, the entanglement entropy, and the subregion complexity are transformed into
\ba
T\to \Lambda_0 T,\quad l\to \Lambda_0^{-1}l, \quad q\to \Lambda_0 q,\quad  S^{EE}\to S^{EE}, \quad C\to C.
\ea
Due to the presence of scaling symmetry, it is convenient to use dimensionless parameters such as $Tl$ and $ql$. 

One needs to use the numerics to compute the holographic subregion complexity. First, we find $z_*$ by following the argument around eq. \eqref{XZ32} and fixing the parameter $l$. Secondly, we obtain $x(z)$ from eq. \eqref{XZ32}  to perform the double integration in eq. \eqref{COM3}. The subregion complexity is divergent itself.  
It can be shown that the divergent part of $C (\epsilon )$ is proportional to only $1/\epsilon$.
 The coefficient of the divergent part is given by
\ba\label{LDI37}
l_d=-\dfrac{\kappa^2\epsilon_1\epsilon_2 (C (\epsilon_1)-C(\epsilon_2))}{\epsilon_1-\epsilon_2}. 
\ea
The parameter $l_d$ should be a function of the size of the interval $l$.~\footnote{In~\cite{Abt:2017pmf}, the $2d$ holographic subregion complexity including the Ricci scalar has the divergent structure $\frac{l}{\epsilon}+a_1$ in an $AdS_3$ black hole by applying the 2-dimensional Gauss-Bonnet theorem. Now, $l$ is the size of the interval and $a_1$ is an Euler number plus a constant. Due to the same theorem, the subregion complexity of the $AdS$ black hole with a vector hair also has the same form in the presence of the Ricci scalar~\eqref{MET1}. Without the Ricci scalar, however, $a_1$ is not topological without the Gauss-Bonnet term. } One needs to subtract this singular part to pick up the finite contribution $\kappa^2 HC_\mathrm{fin}$. 

\begin{figure}[htbp]
     \begin{center}
          \includegraphics[width=2.5in]{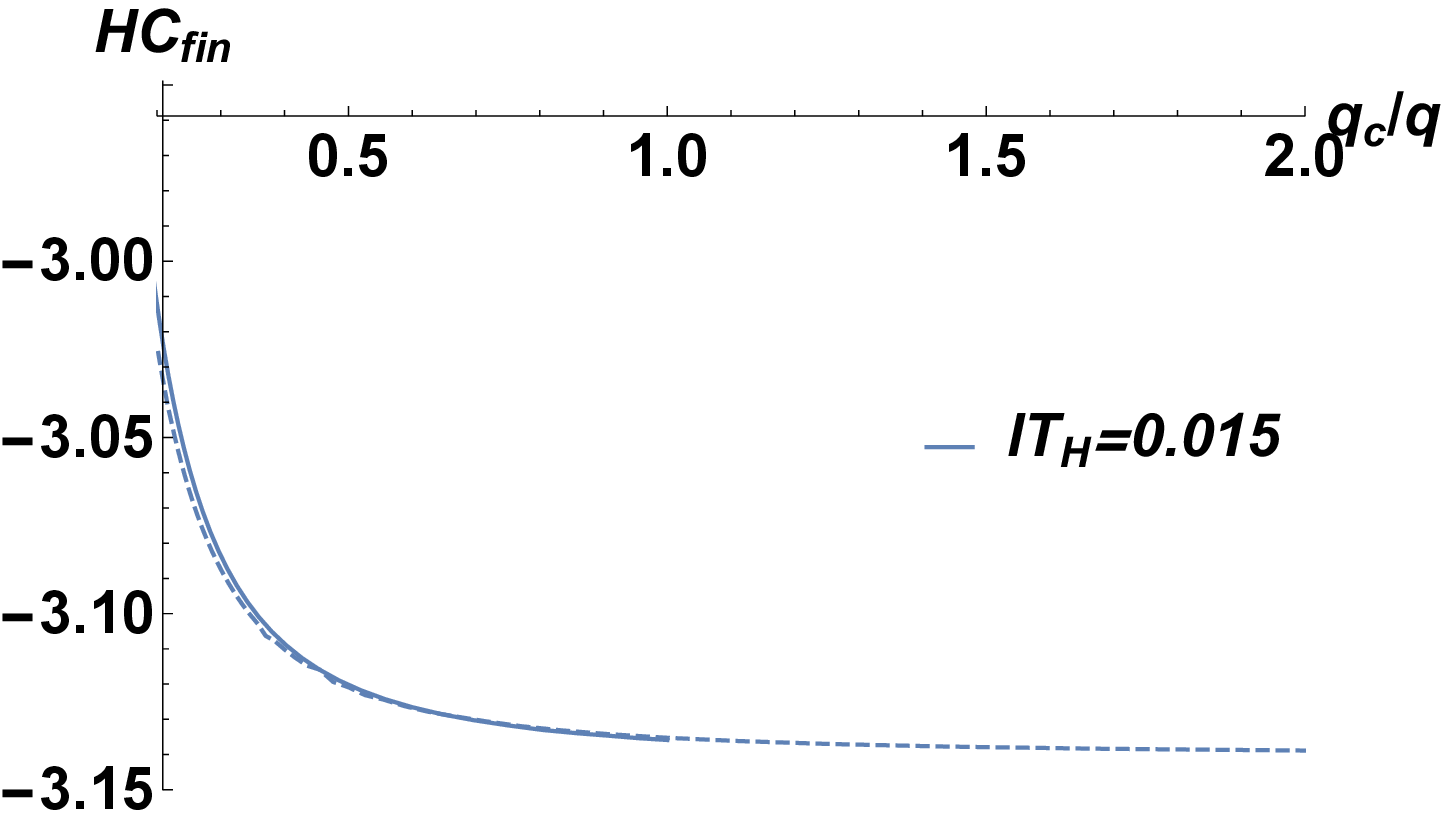} 
          \hspace{1.6cm}
  \includegraphics[width=2.5in]{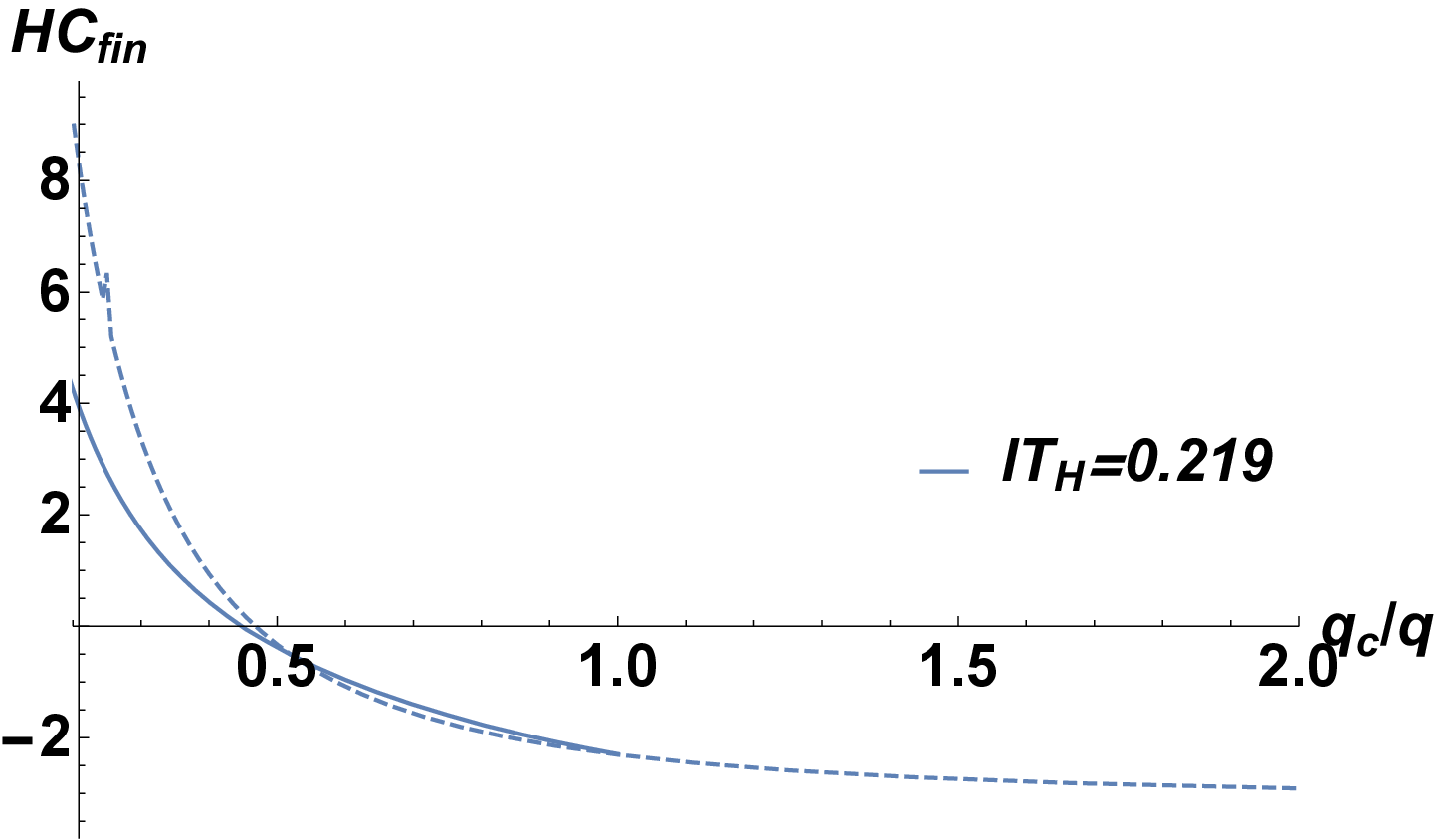} 
           \caption{The normalized finite part  $HC_\mathrm{fin}$ as a function of $q_c/q$ with fixed $T_H$, $l$ and fixed $\tilde{\kappa}^2=0.1$ ($q_c=33.5T_H=5.02$). Left: When $lT_H\ll 1$, the holographic complexity coincides between the normal and condensed phases. Right: Due to the formed condensate, the holographic complexity of the condensed phase turns out to be smaller than the one of the normal phase at high charge density. } 
    \label{fig:hcq}
    \end{center}
     \begin{center}
          \includegraphics[width=2.5in]{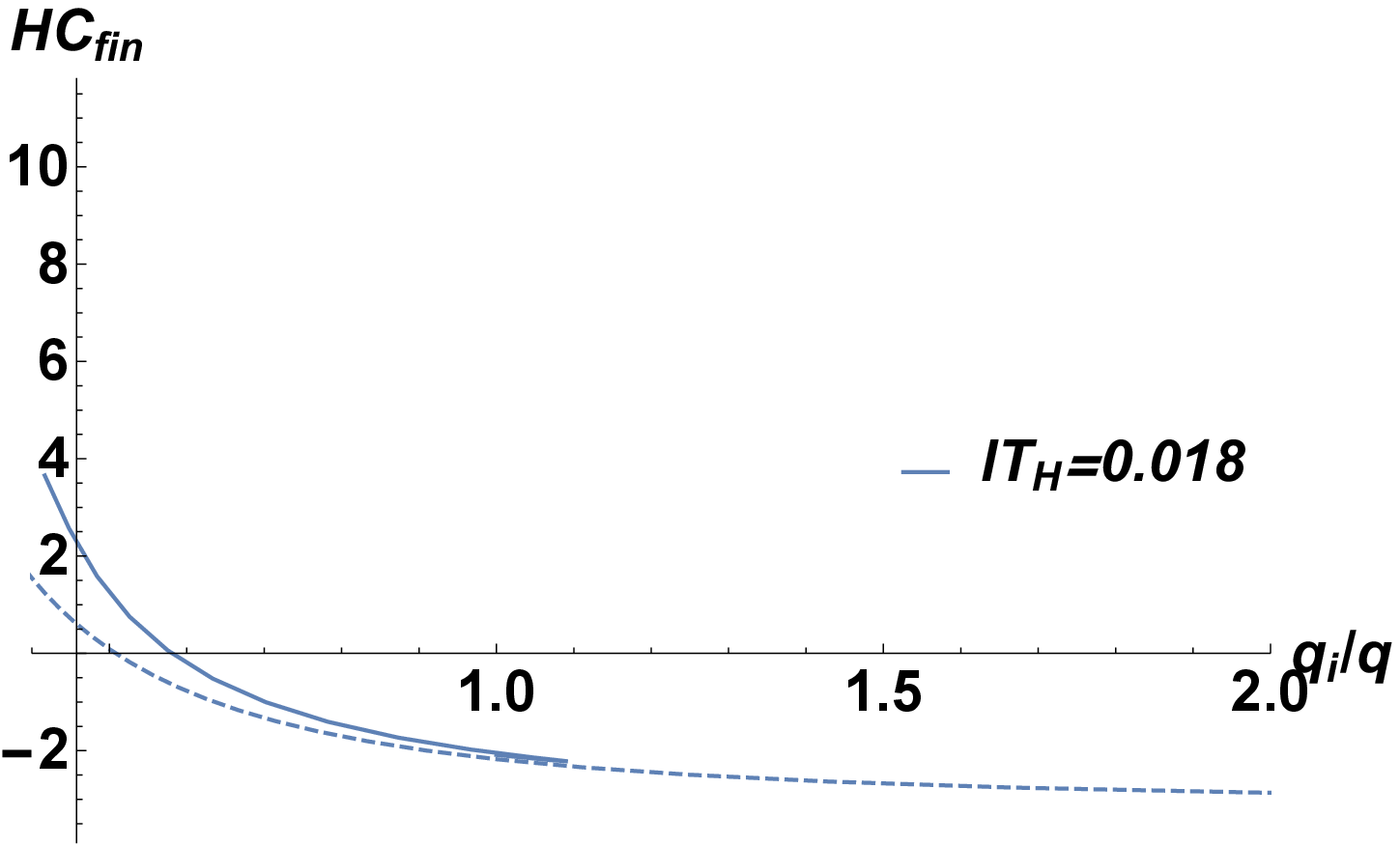} 
          \hspace{1.6cm}
  \includegraphics[width=2.5in]{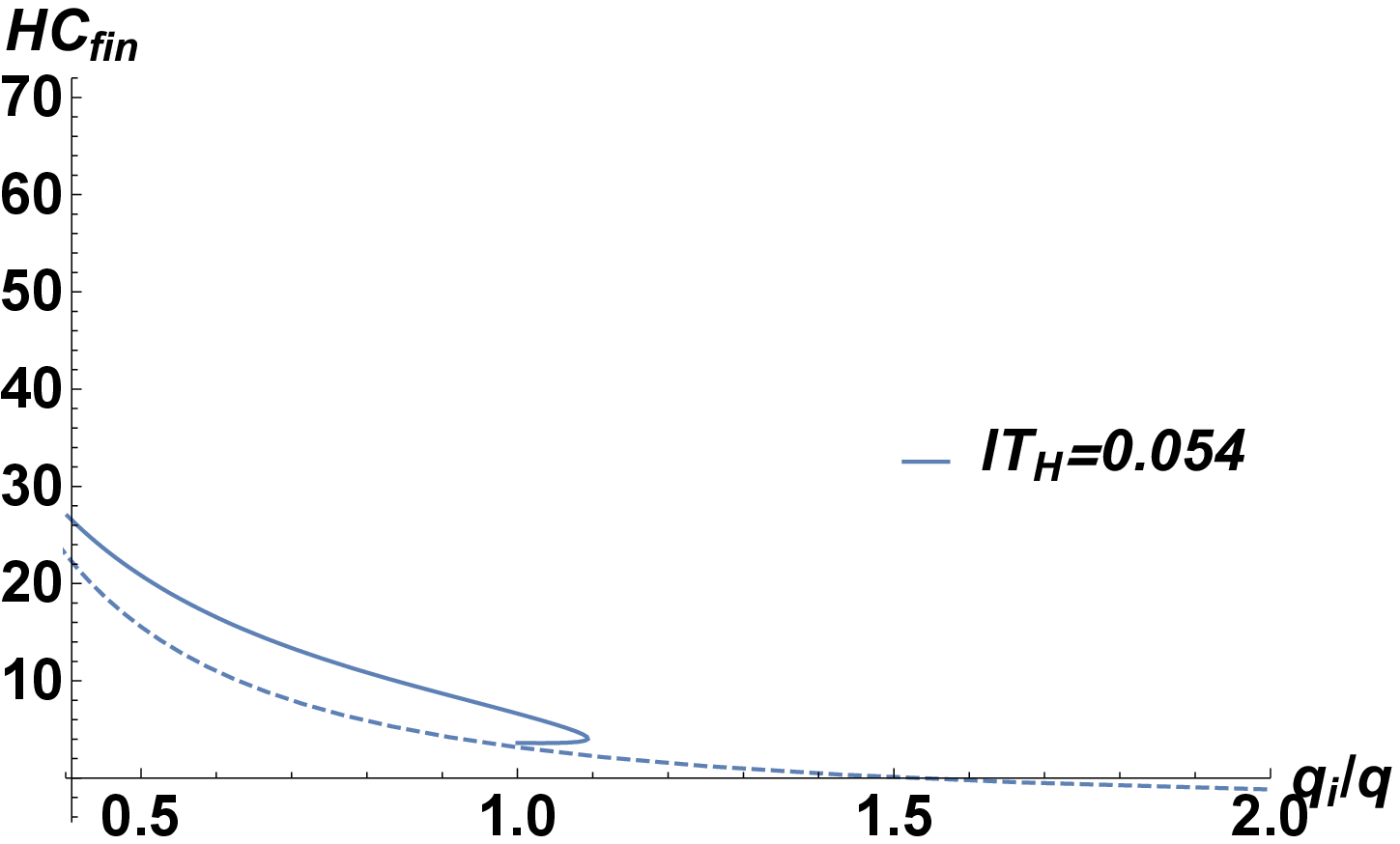} 
           \caption{ The normalized finite part $HC_\mathrm{fin}$ as a function of $q_i/q$ with fixed $T_H$, $l$, and fixed $\tilde{\kappa}^2=0.5$ ($q_c=189T_H=28.3$). In the left figure, $HC_\mathrm{fin}$ of the condensed phase starts from $q=q_i$. It turns out to be multi-valued when the charge density is small $q<q_i$. The open angle $\theta_o$ between two phases is almost $\pi$. In the right figure, the open angle $\theta_o$ is smaller at fixed $lT_H=0.054$. } 
    \label{fig:hcq2}
    \end{center}
\end{figure}
\begin{figure}[htbp]
     \begin{center}
          \includegraphics[width=2.5in]{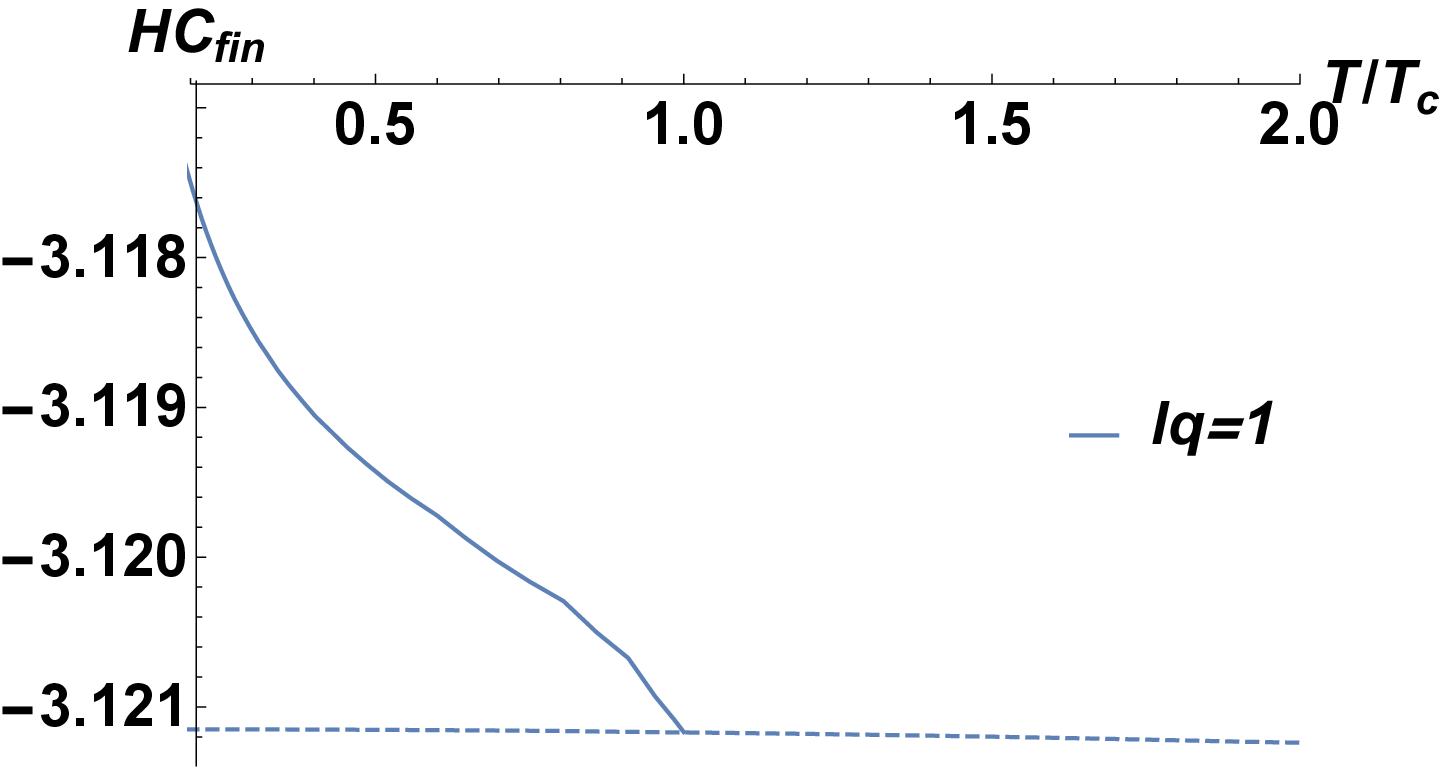} 
          \hspace{1.6cm}
  \includegraphics[width=2.5in]{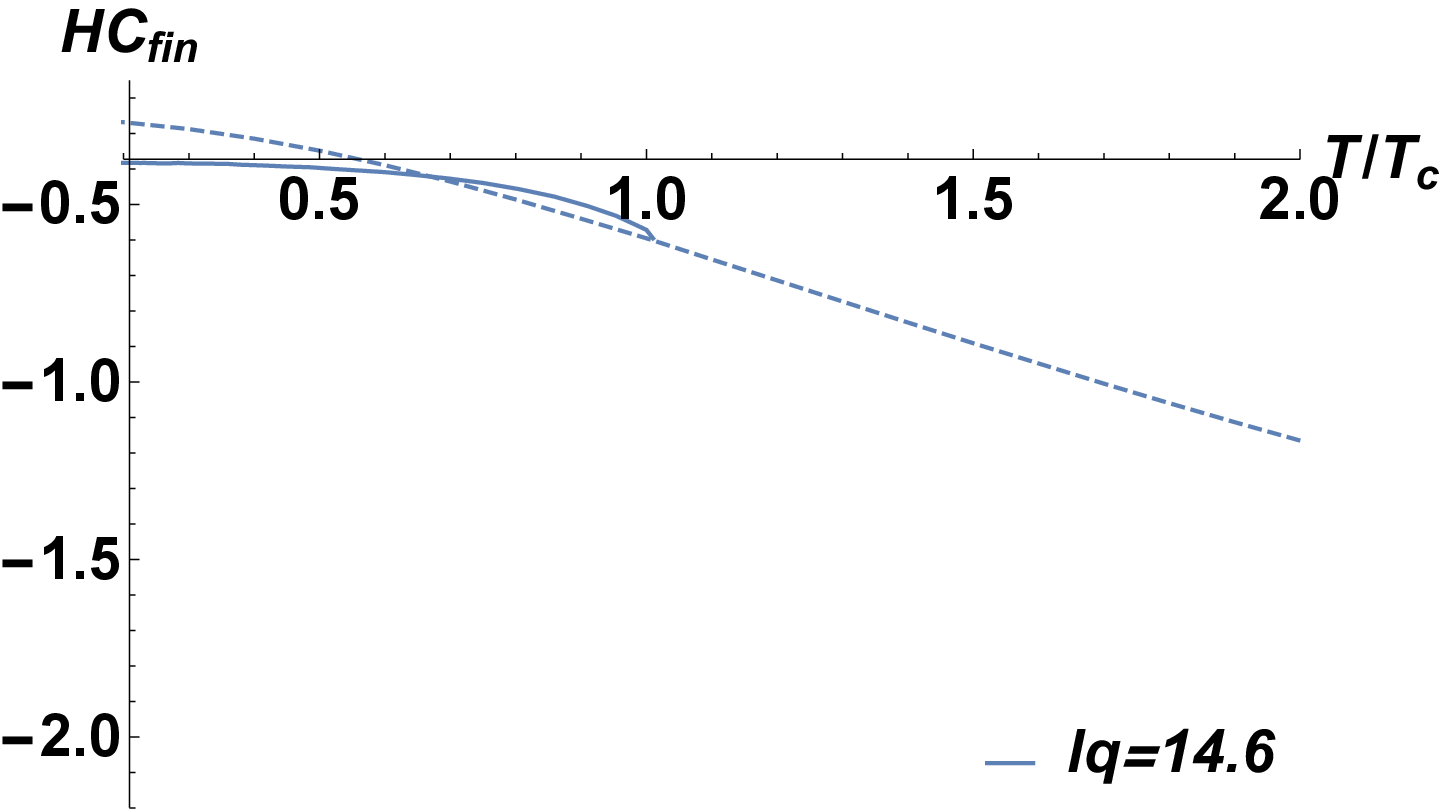} 
         \hspace{1.6cm}
           \caption{The finite part $HC_\mathrm{fin}$ as a function of $T/T_c$ with fixed $q$, $l$, and fixed $\tilde{\kappa}^2=0.1$ ($T_c=0.03q=0.3$). In the normal phase, $HC_\mathrm{fin}$ is always a decreasing function of $T/T_c$. Due to the formed condensate, $HC_\mathrm{fin}$ in the normal phase becomes smaller than that of the normal phase.} 
    \label{fig:hcT}
    \end{center}
    \begin{center}
    \includegraphics[width=2.5in]{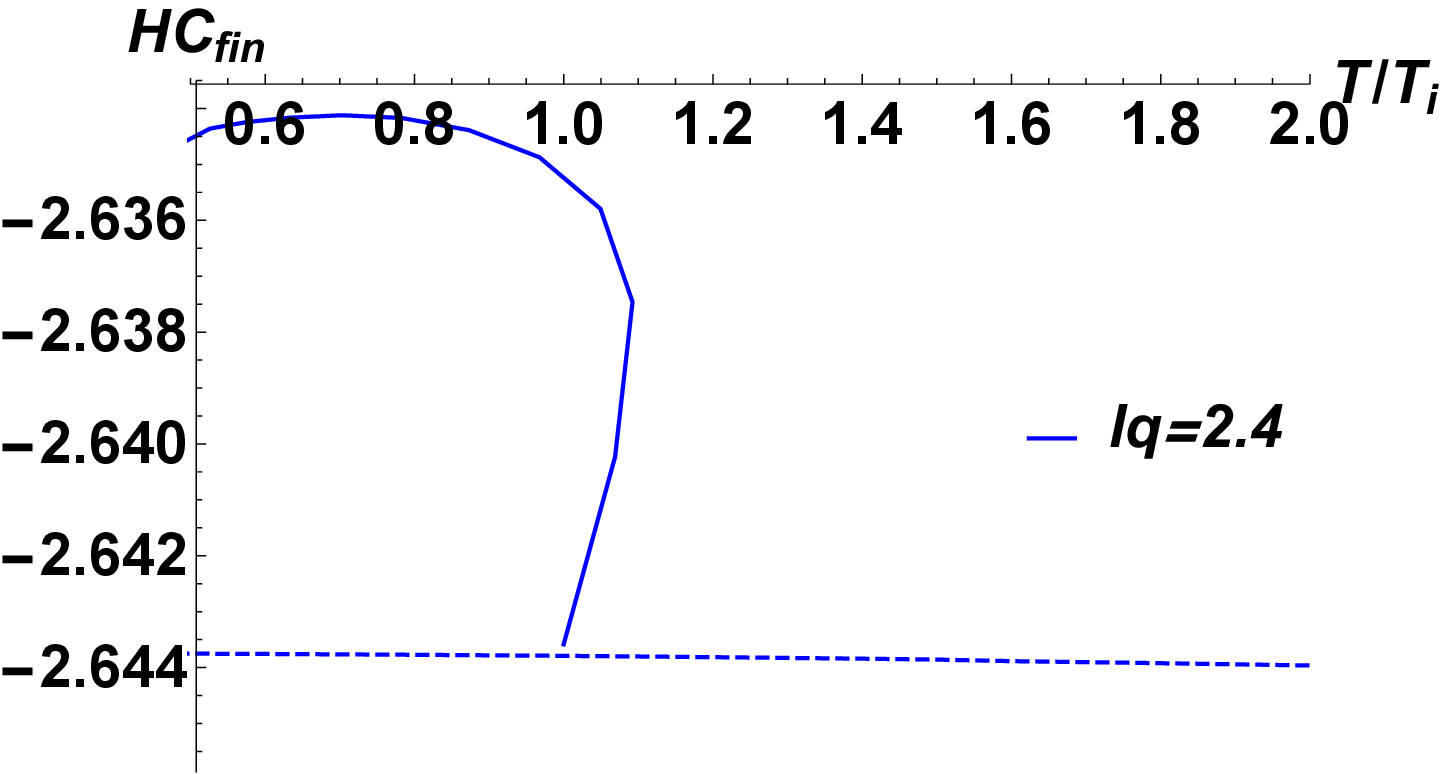} 
\caption{The finite part  $HC_\mathrm{fin}$ as a function of $T/T_c$ with fixed $q$, $l$, and fixed $\tilde{\kappa}^2=0.5$ ($T_c\sim 0.0053 q=0.053$). In the normal phase, $HC_\mathrm{fin}$ decreases with increase of $T/T_c$. By contrast, the holographic complexity in the condensed phase turns out to be multi-valued at a specific range of parameters. It increases with decrease of $T/T_c$ after the phase transition point,   having a peak at low temperature. }
    \label{fig:hcT2}
    \end{center}
\end{figure}

\subsection{The holographic complexity as a function of $q$ or $T$}
We consider two separate coupling constants $\tilde{\kappa}^2=0.1$ and $0.5$ in our numerical computation. We specify the coefficient $l_d$ and the finite part $\kappa^2 HC_\mathrm{fin}$ for each coupling constant. We find that $l_d$ is linearly equal to $l$ in the numerics. This linear behavior of the subregion complexity is also observed in the Ising model on the squared lattice in the context of the tensor network~\cite{Abt:2017pmf}. 

Subtracting the singular part with a coefficient in eq. \eqref{LDI37}, we compute the finite part of the subregion complexity $\kappa^2 HC_\mathrm{fin}$.  We plot the finite part $\kappa^2 HC_\mathrm{fin}$ fixing the size of the interval $l$,  temperature $T_H$, and $\tilde{\kappa}^2$ in Fig. \ref{fig:hcq} and \ref{fig:hcq2}. We plot the finite part $\kappa^2 HC_\mathrm{fin}$ fixing $l$, $q$, and $\tilde{\kappa}^2$ in Fig. \ref{fig:hcT} and \ref{fig:hcT2}. In both cases, when the size $l$ is smaller than $1/T_H$ (the extremal limit) or $1/q$, the finite part $\kappa^2 HC_\mathrm{fin}$ in the condensed phase (solid curves) is almost equal to the one of the normal phase (dashed curves). 
 The two behaves differently when $lT_H\gg 1$ or $lq\gg 1$. Note that the intersecting point arises from $q=q_i(T=T_i)$. While the intersecting point coincides with the critical point $q_i=q_c$ for $\tilde{\kappa}^2=0.1(<0.31)$, $q_i\neq q_c$ for $\tilde{\kappa}^2=0.5(>0.31)$. When  $\tilde{\kappa}^2=0.5(>0.31)$, the intersecting point does not seem to have physical meanings.

In the normal phase of both coupling constants, the finite part $\kappa^2 HC_\mathrm{fin}$ decreases with increase of  $T/T_c$ or $q_c/q$. When  $\tilde{\kappa}^2=0.1$, the finite part $\kappa^2 HC_\mathrm{fin}$ in the condensed phase behaves similarly.  This implies that the ordered phase at high density is a more complicated system. When $\tilde{\kappa}^2=0.5$, the finite part $\kappa^2 HC_\mathrm{fin}$ turns out to be multi-valued at a specific range of the parameter $q_c/q$ or $T/T_c$. As opposed to small $\tilde{\kappa}<0.31$, the finite part $HC_\mathrm{fin}$ in the condensed phase has a peak at the intermediate regime after increasing at low temperature in Fig. \ref{fig:hcT2}. These behaviors are separate from the holographic entanglement entropy, which decreases at low temperature or high density. 

One can define opening angles $\theta_o$ around the intersecting point $q_i$ or $T_I$ between two curves of two phases. The opening angles $\theta_o$ increase when $\tilde{\kappa}^2$ increases as observed in the holographic $s$-wave superconductor~\cite{Zangeneh:2017tub}. When $\tilde{\kappa}^2<0.31$, $\theta_o$ is small, being similar to the probe limit. When $\tilde{\kappa}^2>0.31$, $\theta_o$ can turn out to be larger than $\pi/2$. 

The extremal limit $T_Hl\ll 1$ of opening angles is interesting. In the extremal limit and for $\tilde{\kappa}^2<0.31$, the holographic complexity of the condensed phase coincides with the one of the normal phase. The opening angles between the normal and condensed phases are small enough in the extremal limit. By contrast, while the holographic complexity coincides between two phases for $\tilde{\kappa}^2>0.31$ in the extremal limit, the open angles between two phases are large enough. See the right hand side of Fig. \ref{fig:hcq2}. The open angles decrease with increase of temperature.  

In summary, we analyzed $T$ or $q$ dependence of the subregion complexity. The universal part $HC_u$ is always finite in both phases. When $lq$ or $lT$ is much smaller than 1 (the pure $AdS_3$ limit), two curves almost agree (c.f. the case of the holographic entanglement entropy). By increasing $lq$ or $lT$, the volume surface probes the region near the black hole horizon. As seen in Fig.~\ref{fig:hcq} and Fig. \ref{fig:hcT} for $\tilde{\kappa}^2=0.1$, $HC_u$ in the condensed phase becomes even smaller than the one of the normal phase at low temperature or high density. As seen from the analysis of the holographic entanglement entropy, the condensate dominates charged degrees of freedom at a large subregion. It implies that the condensate decreases $HC_u$ with the small backreaction. When $\tilde{\kappa}$ is large $(\tilde{\kappa}^2=0.5)$, however, a large $l$ was not be able to be chosen due to a numerical problem. 

\subsection{The holographic complexity as a function of $lq$}
In this section, we compute $\kappa^2 HC_\mathrm{fin}$ as well as $S^{EE}_{fin}$ as a function of $lq$. After increasing the size $l$, the minimal surface is attached to the black hole horizon. A part of the surface attached to the black hole horizon explains the thermal entropy. Following~\cite{Azeyanagi:2007bj}, moreover, the minimal surface wraps the black hole horizon of a BTZ black hole for the enough large size of the interval, which gives the contribution of the thermal entropy. Thus, we introduce the following surface 
\ba
S^{EE}=S_{ent}+S^{EE}(2\pi -l),
\ea  
where $S_{ent}$ is thermal entropy. That is, the difference $S^{EE}(2\pi -\delta)-S^{EE}(\delta)$ $(\delta \ll1)$ is equal to the thermal entropy $S_{ent}$. This equality is also satisfied by the entanglement entropy of $2d$ free massless Dirac Fermions  on the 2-torus~\cite{Azeyanagi:2007bj}.

\begin{figure}[htbp]
     \begin{center}
 \includegraphics[width=2.5in]{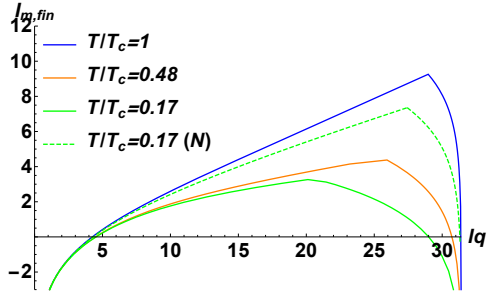} 
\caption{The finite part of the entanglement entropy $\kappa^2 S^{EE}_{fin}/(2\pi)=I_{m,fin}$ as a function of $lq$, where $q=5$.  The critical size of the phase transition turns out to be $l_cq=29.1,\ 25.7, \ 20.5$ for $T/T_c=1,\ 0.48,\ 0.17$, respectively. The dashed green curve is the finite part in the normal phase with the same charge density. The critical size is $l_cq=27.5$. The finite part of the entanglement entropy decreases with decrease of temperature. }
\label{fig:holoe}
\end{center}
\end{figure}

Actually, the surface wrapping the black hole horizon minimizes the holographic entanglement entropy when the size $l$ is larger than a critical size $l_c$. There is a phase transition at a critical size $l_c$ as a function of $l$. In Fig. \ref{fig:holoe}, the finite part of the entanglement entropy is plotted as a function of $lq$, where $q=5$. After varying the size of the interval, there is a phase transition of the entanglement entropy. The critical size of the phase transition turns out to be $l_cq=29.1,\ 25.7, \ 20.5$ for $T/T_c=1,\ 0.48,\ 0.17$ ($T_c=0.03 q=0.149$), respectively. The dashed green curve means the finite part in the normal phase. The critical size becomes $l_cq=27.5$. The critical size decreases with decrease of temperature. The blue curve is almost the same as in the charged $AdS_3$ black hole with same parameters. The finite part of the entanglement entropy becomes small with decrease of temperature.

\begin{figure}[htbp]
     \begin{center}
 \includegraphics[width=2.5in]{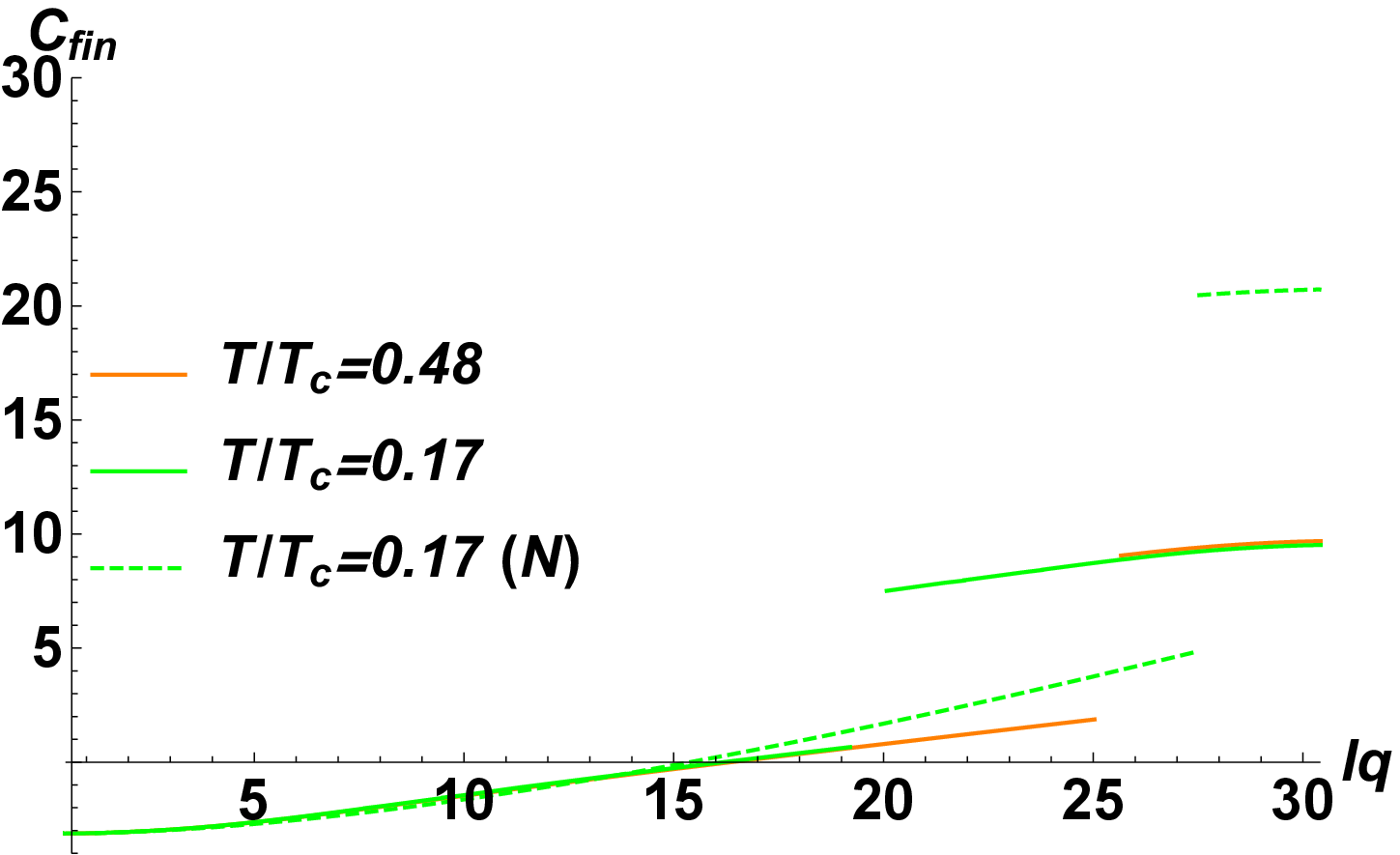} 
          \hspace{1.6cm}
 \includegraphics[width=2.5in]{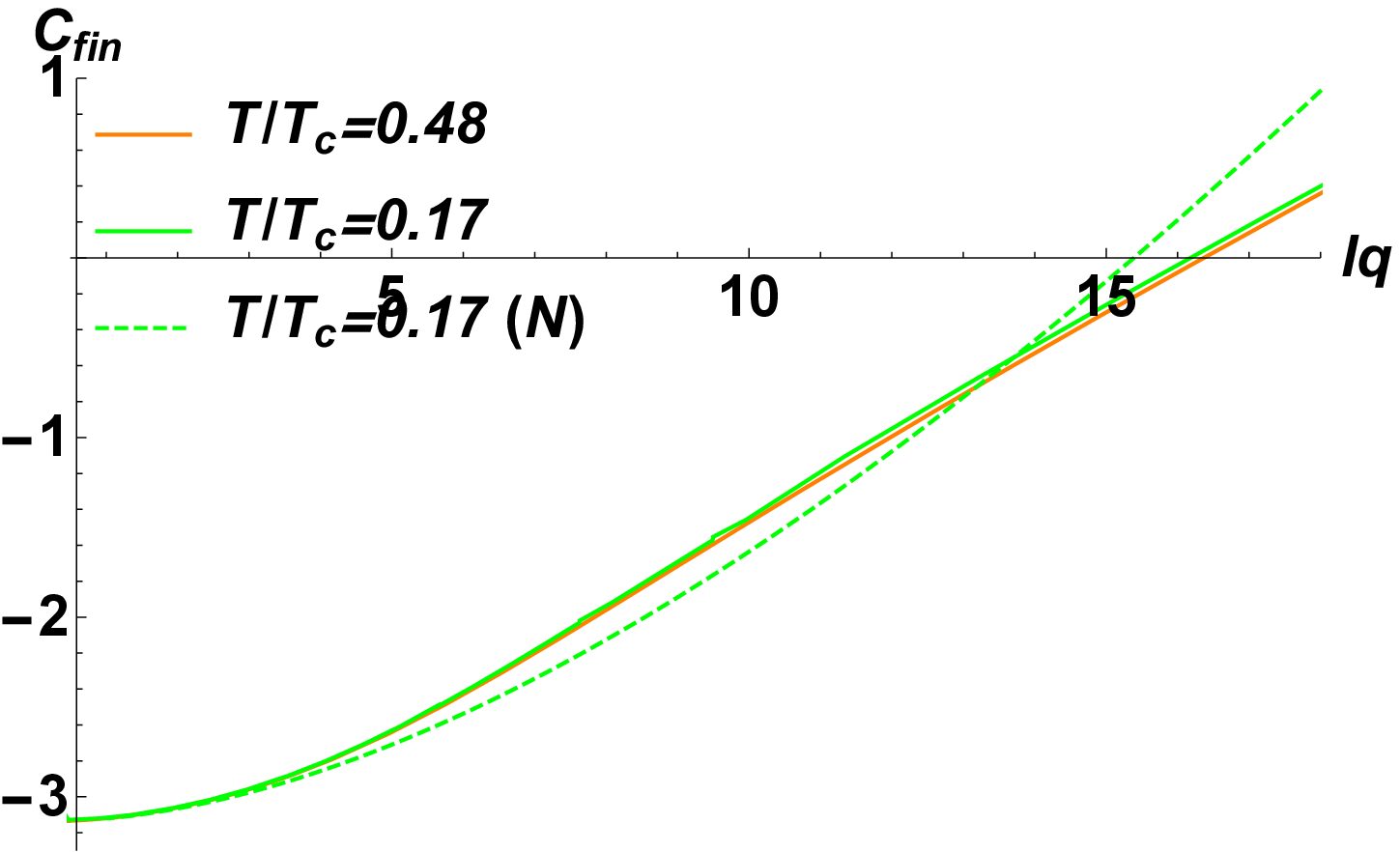} 
           \caption{Left: The normalized finite part $HC_\mathrm{fin}$ suddenly jumps at a critical size $l_cq=25.7,\ 20.5$ ($q=5$) for fixed $T/T_c=0.48,\ 0.17$, respectively. The critical values decrease with decrease of $T/T_c$. The dashed green curve is $HC_\mathrm{fin}$ in the normal phase with the same charge density. The critical size is $l_cq=27.5$. Right: Close-up figure of the left-hand side.  The finite part $HC_\mathrm{fin}$ in the normal phase becomes even larger than that in the condensed phase when $lq$ becomes large.} 
    \label{fig:holoc}
    \end{center}
\end{figure}

Due to the minimal surface wrapping the black hole horizon at a large size $l$, besides, the complexity also has the following form:
\ba\label{COM39}
C=C_{entire}-C(2\pi -l),
\ea
where $C_{entire}$ is the subregion complexity of the entire spatial boundary 
\ba
C_{entire}=\dfrac{c}{3} \int^{z_*}_{\epsilon} dz \dfrac{1}{z^2\sqrt{f(z)h(z)}}.
\ea 
$C_{entire}$ does not give any finite part in an $AdS_3$ black hole. 
In eq. \eqref{COM39}, the singular part is proportional to the size $l$ due to the cancellation between two terms. 
The finite part of the holographic complexity $\kappa^2 HC_\mathrm{fin}$ is plotted as a function of a dimensionless size $lq$ at fixed temperature in Fig. \ref{fig:holoc}. The finite part $\kappa^2 HC_\mathrm{fin}$ increases with increase of the size of the interval. Due to the topological phase transition of the minimal surface surrounding the volume of the subregion complexity at critical sizes,  the finite part $\kappa^2 HC_\mathrm{fin}$ suddenly jumps at a critical size $l_cq=25.7,\ 20.5$ ($q=5$) for $T/T_c=0.48,\ 0.17$, respectively. The dashed green curve is $HC_\mathrm{fin}$ in the normal phase. The critical size is $l_cq=27.5$. The critical size $l_c$ is not dependent on the magnitude of the subregion complexity but the magnitude of the holographic entanglement entropy. The critical size decreases with decrease of temperature. $lq=10\pi$ corresponds to the entire spatial boundary. In the figure, the finite part turns out to be maximum at $lq=10\pi$. The finite part $HC_\mathrm{fin}$ in the normal phase becomes even larger than that in the condensed phase when $lq$ becomes large. Recall that the formed condensate dominates the charged degrees of freedom in the large separation. Because the volume surface at a large size probes the black hole horizon as observed in the holographic entanglement entropy, it implies that the formed condensate decreases the subregion complexity.

The discontinuous jump of the subregion complexity was originally found in \cite{Ben-Ami:2016qex,Abt:2017pmf} by computing topologically different configurations of $2d$ volume surfaces (i.e. the disc as well as the annulus). In an $AdS_3$ black hole, this finite part at zero charge density approaches $\pi$ after the jump from $-\pi$ to the tolopogically different configuration. That is the opposite sign of the finite part. The magnitude of the jump is $\Delta C=2\pi$ being independent of temperature. After switching on the charge density, the difference $\Delta HC_\mathrm{fin}(l=l_c)$ is not independent of the charge density but almost $2\pi$ in the condensed phase, while the jump is larger than $2\pi$ in the normal phase in Fig. \ref{fig:holoc}. 

In summary, the singular part of the subregion complexity gives an expected linear behavior, which is divergent like $l/\epsilon$ for those in either eq. \eqref{COM3} or eq. \eqref{COM39}. The subregion complexity jumps at the critical length. Depending on charge density, the difference $\Delta HC_\mathrm{fin}(l=l_c)$ changes more in the normal phase.  Interestingly, the maximum of the subregion complexity is not a constant depending on charge density $q$ or temperature $T$. As shown in Fig. \ref{fig:holoc}, $HC_\mathrm{fin}$ in the condensed phase can become smaller than the one in the normal phase. Except for the discontinuous phase transition, the formation of the condensate dominates the charged degrees of freedom and decreases $HC_\mathrm{fin}$ at the large length.  

\section{The holographic renormalized entanglement entropy}\label{REE}
\begin{figure}[htbp]
     \begin{center}
          \includegraphics[width=2.5in]{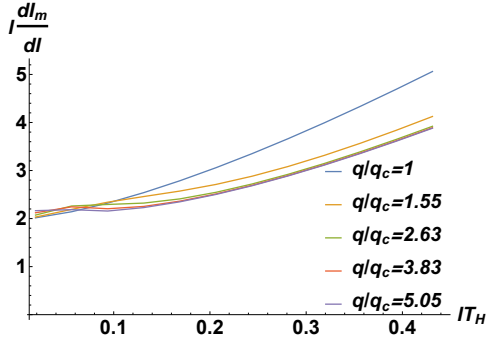} 
           \caption{The renormalized EE $\kappa^2 S^{ren}/2\pi (=l\partial I_m/\partial l)$ as a function of $l$ at fixed temperature $T_H=0.15$ ($\tilde{\kappa}^2=0.1$). Each curve describes the renormalized EE at fixed charges. It obeys the crossover like the one of an $AdS_3$ black hole. The renormalized EE becomes small at large $q$ and $l$, where the formed condensate dominates charged degrees of freedom.} 
    \label{fig:ren1}
    \end{center}
\end{figure}

\begin{figure}[htbp]
     \begin{center}
          \includegraphics[width=2.5in]{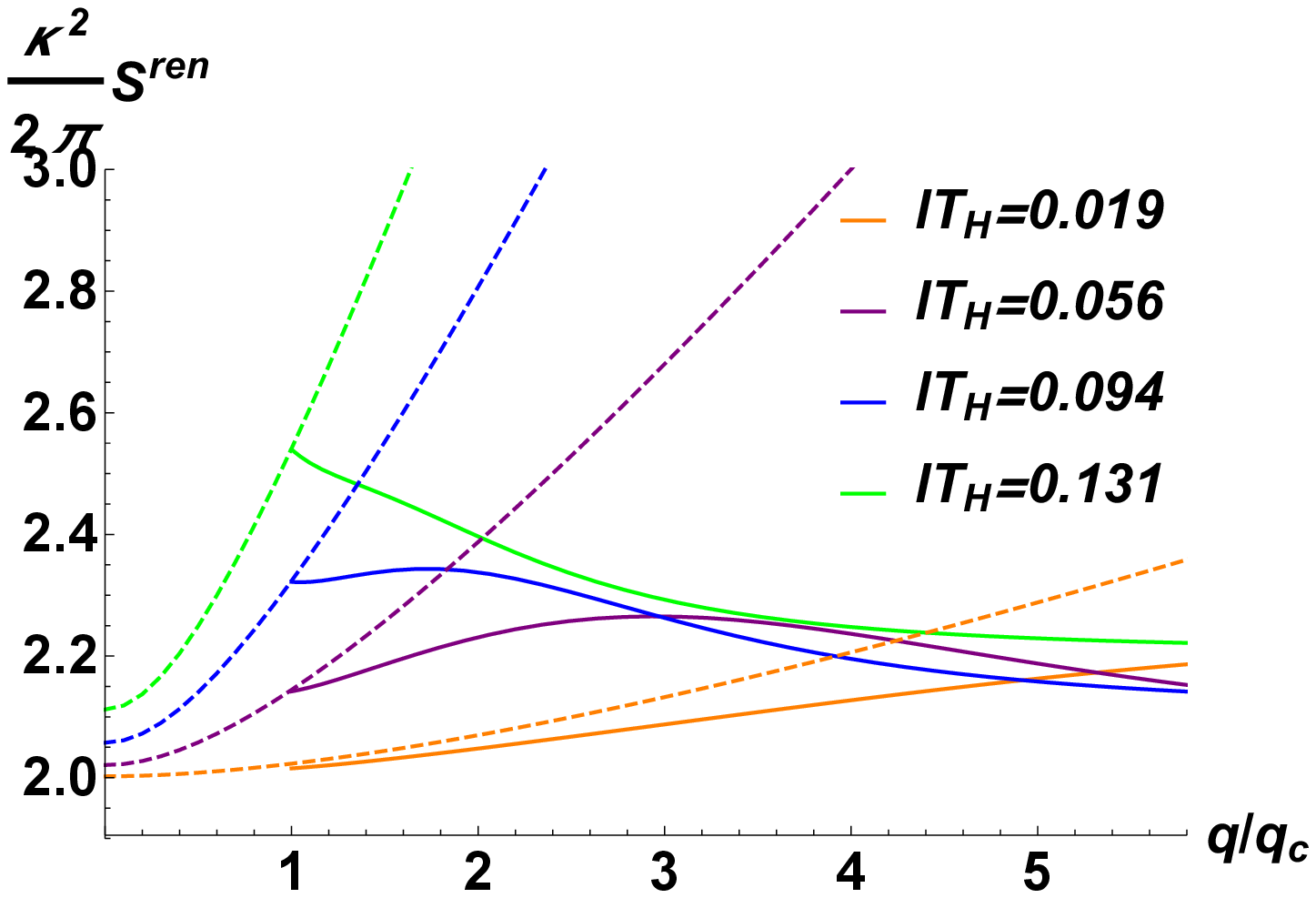} 
          \hspace{1.6cm}
  \includegraphics[width=2.5in]{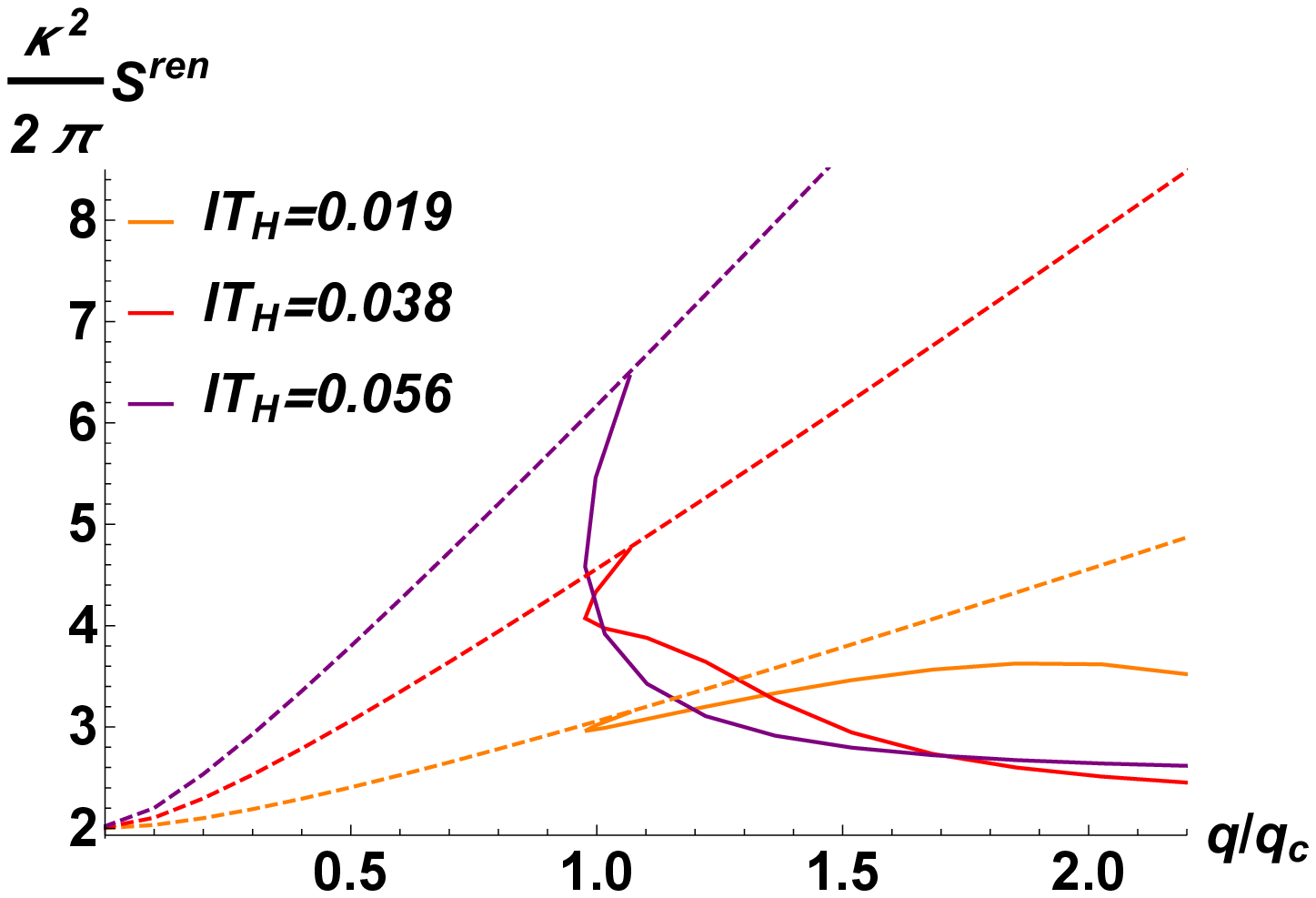} 
           \caption{The renormalized entanglement entropy as a function of $q/q_c$. Left: $\tilde{\kappa}^2=0.1$ Right: $\tilde{\kappa}^2=0.5$. In both cases, the renormalized entanglement entropy decreases at a large size $l$ and charge density $q$. } 
    \label{fig:ren3}
    \end{center}
\end{figure}

Motivated by the analysis of the universal term in the entanglement entropy, we compute the renormalized entanglement entropy (renormalized EE). The renormalized entanglement entropy is defined as $S^{ren}=l\partial S_{1}/dl$ in 2-dimension~\cite{Casini:2004bw}, where $S_1$ is the entanglement entropy. It becomes UV finite and is independent of the cutoff. Due to the UV finiteness, the renormalized EE is considered as a universal term in the entanglement entropy. This entropy is related with the degrees of freedom at the scale of the length $l$. Moreover, the renormalized EE satisfies the C-theorem only in special cases. 

Hereby, we consider the 2-dimensional thermal CFT at finite temperature. Substituting the entanglement entropy $S_1=\frac{c}{3}\log \Big(\frac{\beta}{\epsilon \pi}\sinh \Big(\frac{\pi l}{\beta}\Big)\Big)$, the renormalized entanglement entropy becomes
\ba\label{SRE22}
&S^{ren}=\dfrac{c\pi l}{3\beta}\coth \Big(\dfrac{\pi l}{\beta}\Big) \nonumber \\
&\sim\begin{cases} & \dfrac{c}{3} \quad (lT\ll 1), \\
&\dfrac{c\pi l}{3\beta} \quad (lT \gg 1) ,
\end{cases}
\ea
 In the small $lT$ limit, moreover, it approaches the vacuum behavior, which does not depend on $l$. In the large $lT$ limit, it approaches the thermal entropy with a positive coefficient times $l$~\cite{Liu:2012eea}. This behaivor is called a crossover.

We apply the renormalized entanglement entropy (renormalized EE) to a holographic $1+1d$ $p$-wave superconductor. The renormalized EE in a 3-dimensional gravity is defined as
\ba 
S^{ren}=l\dfrac{\partial S^{EE}}{\partial l},
\ea
where $S^{EE}$ is Ryu-Takayanagi formula $S^{EE}=\frac{2\pi}{\kappa^2}(\gamma_A)$. The renormalized EE of an $AdS_3$ black hole is given by \eqref{SRE22} when the minimal surface does not wrap the black hole horizon.

In Fig. \ref{fig:ren1}, the renormalized EE is plotted as a function of $l$ with a fixed $q$. When $l$ approaches zero, the renormalized EE goes to a constant, while it is proportional to the thermal entropy density at large $l$. So, it obeys the crossover like the one of an $AdS_3$ black hole. The renormalized EE in the condensed phase becomes smaller than the critical behavior $q=q_c$ at large $l$. This implies that the formation of the condensate dominates the charged degrees of freedom by probing the black hole horizon.
%%As the charge $q$ increases, the renormalized EE has extremal values. 
%%It behaves non-monotonically as a function of $l$.

In Fig. \ref{fig:ren3}, the renormalized EE is plotted as a function of $q/q_c$ when $l$ is fixed (left: $\tilde{\kappa}^2=0.1$ and right: $\tilde{\kappa}^2=0.5$). The renormalized EE behaves qualitatively similar to the finite part of the holographic entanglement entropy. When $q>q_c$ and the backreaction is small $\tilde{\kappa}^2=0.1$, it increases at small $lT_H$ and decreases at large $lT_H$. The renormalized EE has a non-monotonic behavior and  extremal values among a range of the size $l$. It implies that there are competing contributions between the charged degrees of freedom and the formed condensate. By probing the black hole horizon, both quantities capture the physics of the formed condensate.

\section{Discussion}
We computed the holographic subregion complexity in a fully backreacted metric of the $1+1$ dimensional $p$-wave superconductor phase transition. We computed the subregion complexity by fixing $q$ or $T$ (or both quantities). We confirm that the universal part $HC_u$ is finite across the phase transition and has competitive behaviors different from the finite part of entanglement entropy as seen in Fig. \ref{fig:holoe} and Fig. \ref{fig:holoc}. We probed the black hole with a vector hair by  changing the size of the subregion complexity. By increasing the size $lq$ or $lT$, the volume surface of the subregion complexity approaches the black hole horizon. As observed in Fig.~\ref{fig:hcq} and Fig. \ref{fig:hcT} for $\tilde{\kappa}^2=0.1$, $HC_u$ in the condensed phase is even smaller than the one in the normal phase at low temperature. As seen from the analysis of the holographic entanglement entropy, the formed condensate dominates the charged degrees of freedom in the large size. It implies that the formed condensate decreases $HC_u$.

The leading divergence of the subregion complexity was shown to be linear to the size of the interval  $C\propto l/\epsilon$ in either eq. \eqref{COM3} or eq. \eqref{COM39}. Even if Ryu-Takayanagi surface wraps the black hole horizon at the large size of the interval, cancellations occur between two terms in eq. \eqref{COM39}. The same linear behavior was observed in the Ising model on the square lattice. 

The finite part of the subregion complexity was plotted by fixing the size of the interval $l$. In the extremal limit $lT_H\ll 1$ or $lq\ll 1$, $\kappa^2 HC_\mathrm{fin}$ almost agreed between normal and condensed phases except for the region around the intersecting point, where the curve in the condensed phase ended.  The curve in the condensed phase behaved differently when $lT_H\gg 1$ (or $lq\gg 1$). Moreover, it depended on the coupling constant. When $\tilde{\kappa}^2=0.1<0.31$, $\kappa^2 HC_\mathrm{fin}$ decreased with increase of $T/T_c$ or $q_c/q$.  This implies that the system at high charge density and low temperature is complicated. This result of the subregion complexity in addition to the finiteness of the universal part agreed with those of the holographic $1+1$ dimensional $s$-wave superconductor~\cite{Zangeneh:2017tub}.

The order of phase transition is varied in the holographic $1+1$ dimensional $p$-wave superconductor with the large amount of the backreaction $\tilde{\kappa}^2=0.5>0.31$, while it is not varied in the holographic $1+1$ dimensional $s$-wave superconductor~\cite{Ren:2010ha,Liu:2011fy,Momeni:2016ekm} with the backreaction. Due to the large amount of the backreaction, the condensate does not behave as in mean field theories (2nd order phase transition) but suddenly jumps to a finite value at the critical point. This large amount of the backreaction also causes the non-monotonic behavior of the finite part $\kappa^2 HC_\mathrm{fin}$ in the condensed phase, while $\kappa^2 HC_\mathrm{fin}$ in the normal phase behaves monotonically. Moreover, the finite part turns out to be multi-valued as a function of $T/T_c$ or $q_c/q$.

We plotted the finite part of the subregion complexity $\kappa^2 HC_\mathrm{fin}$ as a function of $lq$ fixing $q$ and $T$. The formation of the condensate did not almost vary the finite part  $\kappa^2 HC_\mathrm{fin}$, while the charge density varies it. Wrapping the almost entire space circle maximized the subregion complexity. 
We found the discontinuous jump of the finite part $\kappa^2 HC_\mathrm{fin}$. The magnitude of the discontinuous jump depended on the charge density unlike the one of the $AdS_3$ black hole. Note that the magnitude of the jump is larger than the jump in the Ising model on the squared lattice $\Delta HC_\mathrm{fin}\sim 4\pm 0.3 < 2\pi$~\cite{Abt:2017pmf}. This discrepancy will come from broken rotational symmetry of the squared lattice as well as difference between two models. 

Finally, we computed the renormalized EE, which was considered as a universal term of the entanglement entropy. We showed that the renormalized EE had the behavior similar to the finite part of the holographic entanglement entropy: when the backreaction was small, it had a monotonic behavior at both small and large $l$ limits. On the other hand, the renormalized EE behaved non-monotonically for an intermediate region of the size $l$. These can be understood as competition between charged degrees of freedom and the formed condensate. We found that the renormalized EE obeyed a crossover as seen in that of an $AdS_3$ black hole: it approaches a constant for very small size and is linearly proportional to $l$ for a large interval. We noticed that the renormalized EE and  $\kappa^2 HC_\mathrm{fin}$ had some common properties such as decrease due to the formed condensate, approaching a constant in the very small $lq$ limit. We did not compare both quantities in the large size limit due to the presence of the discontinuous phase transition. To compare the renormalized EE with the subregion complexity, it will be interesting to explore the large size limit in higher dimensional holographic models.

\section*{Acknowledgement:}
Special thanks to J. Sun for collaboration during early stages of this work. We would like to thank A. Gadde, B. S. Kim, J. H. Lee, S. Lin, and S. Pujari for helpful discussions and comments. Discussions during the workshop in  Fudan University, “String Theory and Quantum Field Theory,” were useful to complete this work.

\appendix

\section{The free energy}\label{FRE}

\begin{figure}[htbp]
     \begin{center}
          \includegraphics[width=2.5in]{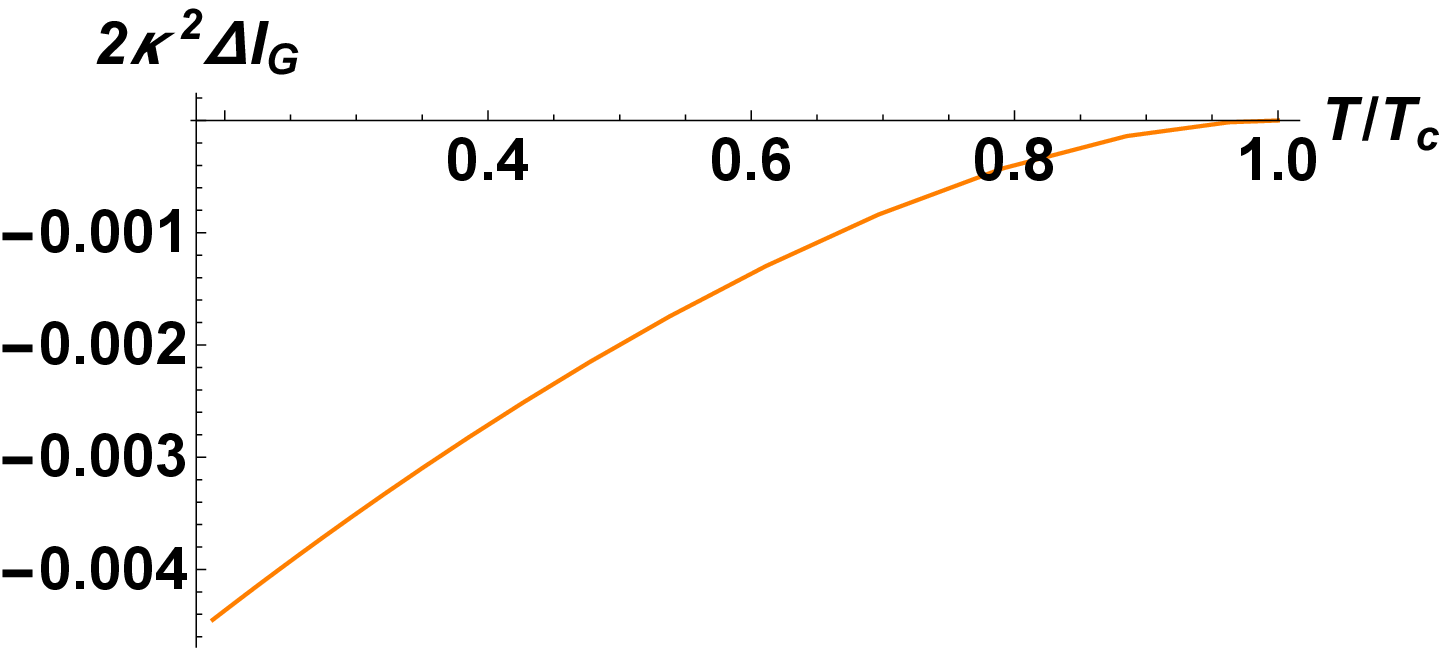} 
          \hspace{1.6cm}
  \includegraphics[width=2.5in]{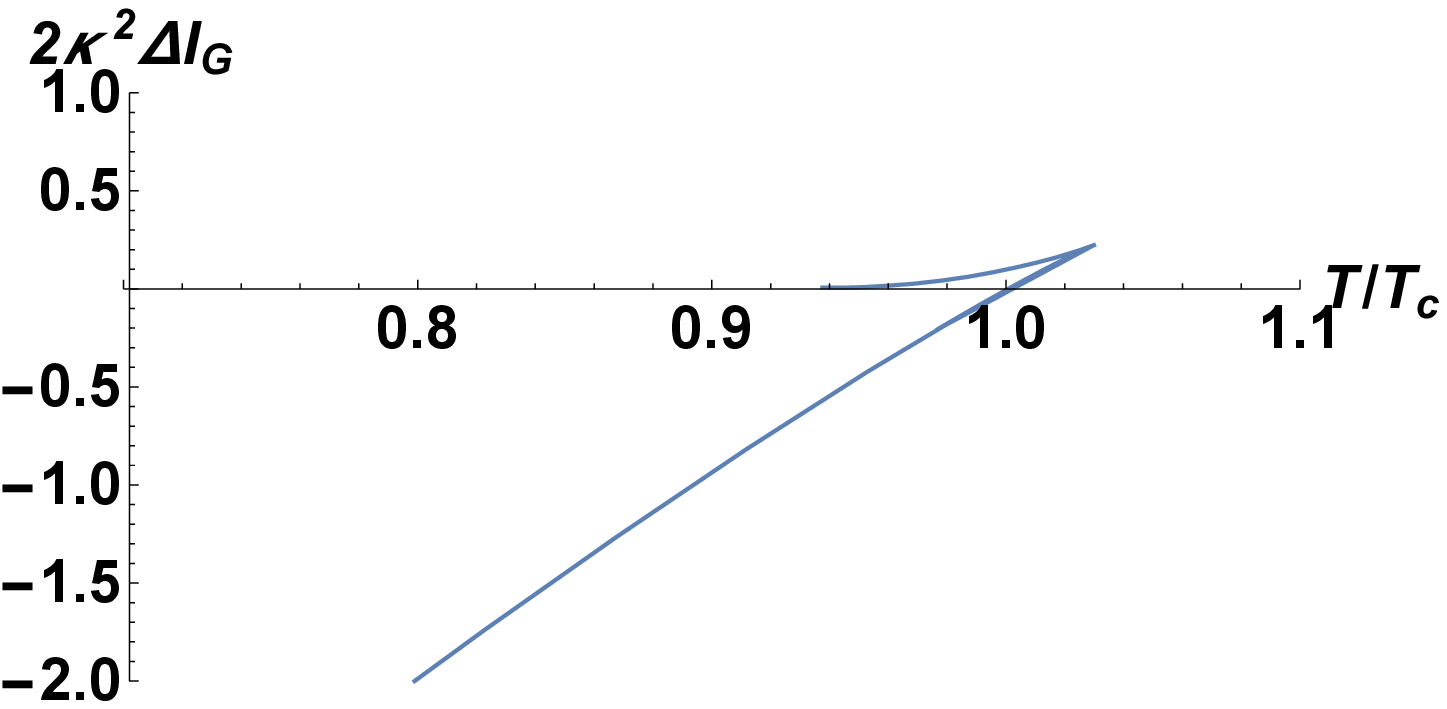} 
           \caption{The difference of normalized free energy $2\kappa^2\Delta I_G$ is plotted at fixed charge density: Left: fixed $\tilde{\kappa}^2=0.1$ and $q=0.5$. The negative value denotes that the condensed phase is favored at temperature lower than $T_c(=0.03 q=0.0149)$. Right: fixed $\tilde{\kappa}^2=0.5$ and $q=30.2$. $\Delta I_G$ is multi-valued at some regime of $T>T_c(\sim 0.0053q=0.16)$. There is the swallow tail of first order phase transition. } 
    \label{fig:freeen}
    \end{center}
\end{figure}
\begin{figure}[htbp]
     \begin{center}
          \includegraphics[width=2.5in]{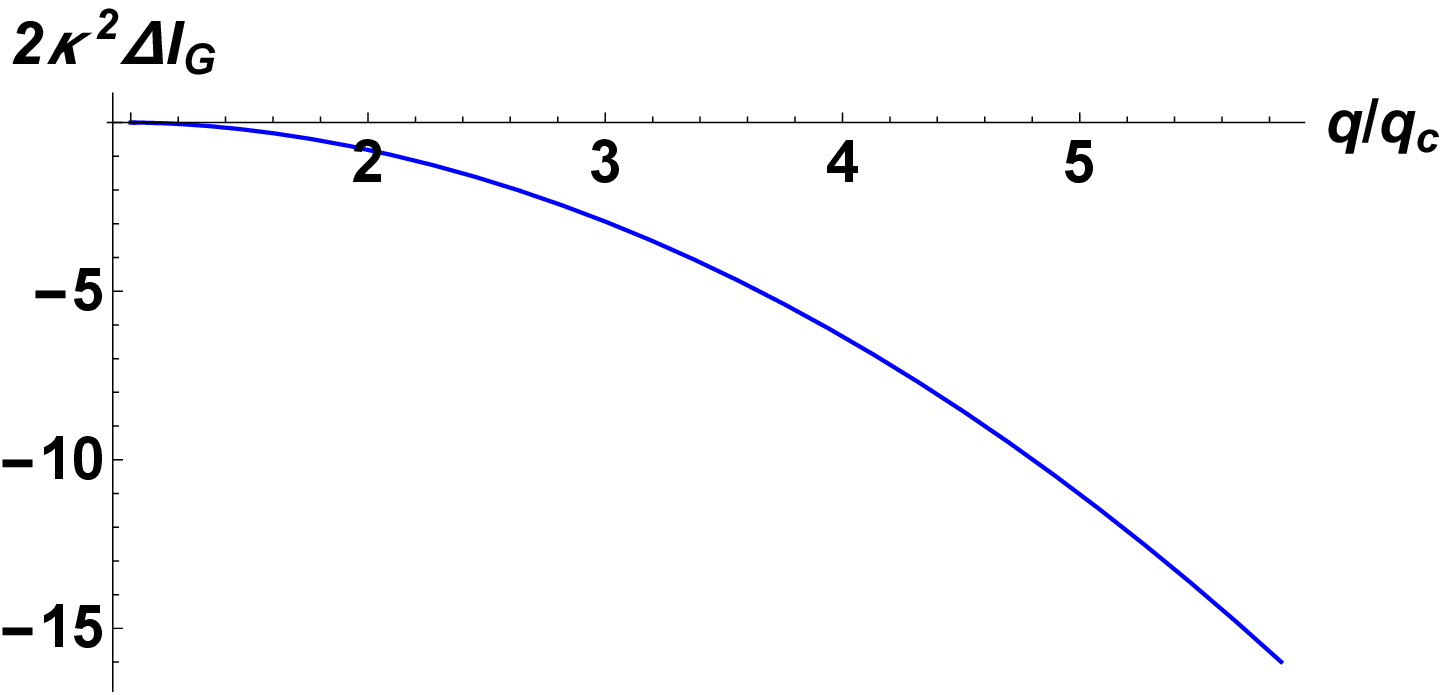} 
          \hspace{1.6cm}
  \includegraphics[width=2.5in]{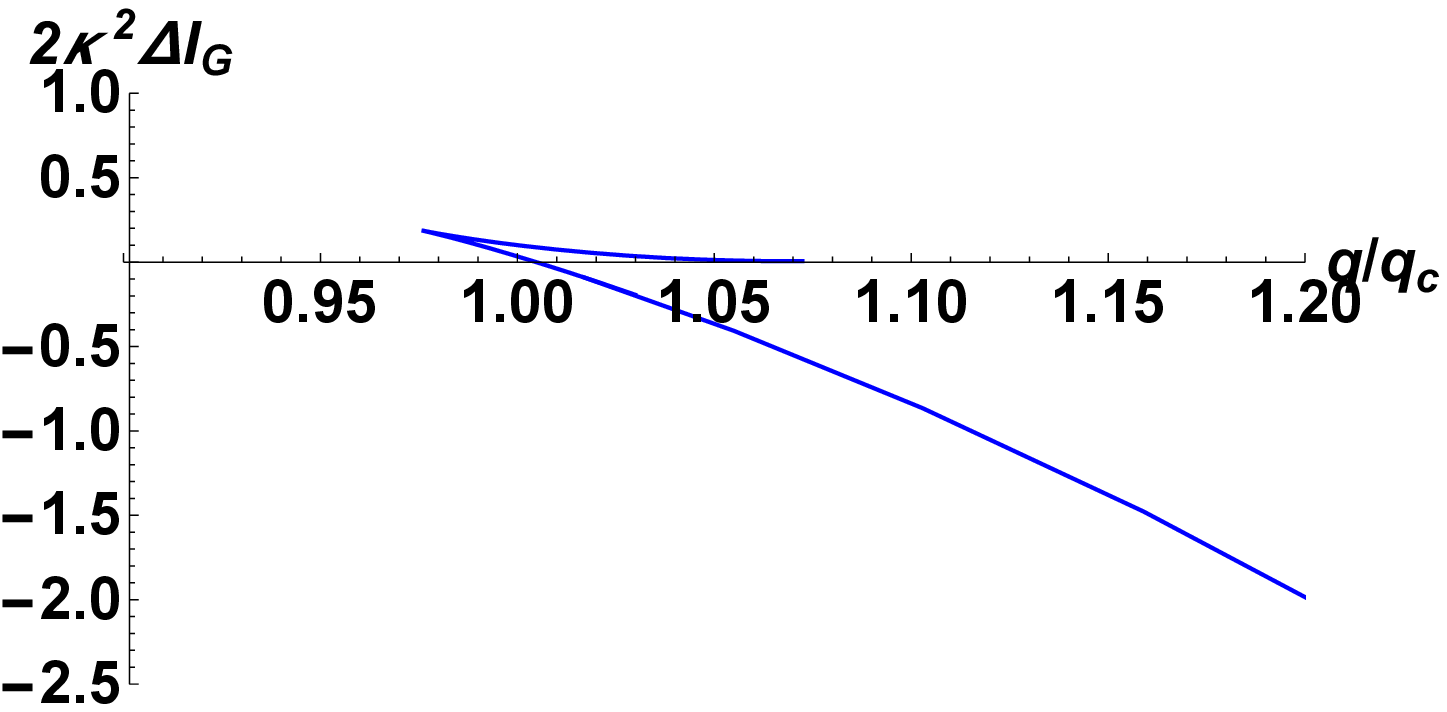} 
           \caption{The difference of normalized free energy $2\kappa^2\Delta I_G$ is plotted at fixed temperature $T_H=0.15$. Left:  fixed $\tilde{\kappa}^2=0.1$. It shows that, at large charge density $q>q_c(=33.5T_H=5.02)$, the condensed phase is favored. Right: fixed $\tilde{\kappa}^2=0.5$. There is the swallow tail of the first order phase transition at the critical charge density $q=q_c(=189T_H=28.3)$. } 
    \label{fig:freeen2}
    \end{center}
\end{figure}

To compute the free energy using the AdS/CFT correspondence, we analyze a finite on-shell action in the presence of the Gibbons-Hawking term and a term of the Legendre transformation. These terms reflect a well-defined variation principle. Due to the divergent on-shell action, we also must use counter-terms to cancel divergence~\cite{Henningson:1998gx,de Haro:2000xn,Karch:2005ms}
\ba\label{A1}
&I_K=\dfrac{1}{\kappa^2}\int d^2x\sqrt{-\gamma}\Big( K+\alpha_1 \Big), \nonumber \\ 
&I_A=\dfrac{1}{ g_{YM}^2}\int d^2x\ \mbox{tr}\Big(-2\sqrt{-g}A_iF^{zi}+\alpha_2 \log \Big(\dfrac{\epsilon}{L}\Big)\sqrt{-\gamma}F_{iz}F^{iz}\Big),
\ea
where $\alpha_1=-1/L$ and $\alpha_2=L$. The log term in the second line addes a scale $L$ in the lagrangian.  
Summing three contributions eq. \eqref{ACT11} and eq. \eqref{A1} up, one can obtain the finite renormalized action and the free energy as follows:
\ba
&I_{tot}=I_G+I_K+I_A, \nonumber \\
&F=-\dfrac{I_{tot}}{\beta},
\ea
where a dictionary of the AdS/CFT correspondence has been used in the last line.

By using the analytic solution of the charged $AdS_3$ black hole eq. \eqref{btz230}, we can integrate the free energy in the normal phase to give
\ba
F=\dfrac{V_1}{\beta}\Big(-\dfrac{L}{2\kappa^2z_h^2}+\dfrac{q^2(1+\log (\frac{z_h}{L}))}{2g_{YM}^2L}\Big),
\ea
where $V_1$ is the volume of the entire spatial circle. In the above free energy, the presence of the log term varies the scaling transformation: the scaling transformation gives an additional term in the free energy~\cite{Jensen:2010em}. 
The variation of the free energy in terms of $T,\ q$ gives thermodynamic quantities such as the entropy $S=-dF/dT$ and the chemical potential $\mu=dF/dq$, respectively. 

 In contrast, the numerical computation is required to compute the free energy in the condensed phase. The difference of the normalized free energy $2\kappa^2\Delta I_G\equiv 2\kappa^2 (F_\mathrm{SF}-F_\mathrm{n})$ is plotted as a function of the normalized temperature $T/T_c$ in Fig. \ref{fig:freeen}. For small $\tilde{\kappa}$ ($\tilde{\kappa}^2<0.31$) and fixed charge density, the solution of the condensed phase always has lower free energy. When $\tilde{\kappa}$ is large $(\tilde{\kappa}^2>0.31)$, $\Delta I_G$ is multi-valued at temperature larger than $T_c$. There occurs the swallow tail of the first order phase transition. In contrast, it is plotted as a function of the normalized charge density in Fig. \ref{fig:freeen2}. While the phase transition to the condensed phase occurs at low temperature $T<T_c$, it occurs at large charge density $q>q_c$. 

\section{The holographic entanglement entropy in the $1+1$ dimensional $p$-wave superconductor}\label{HEEp}
\begin{figure}[htbp]
     \begin{center}
          \includegraphics[width=2.5in]{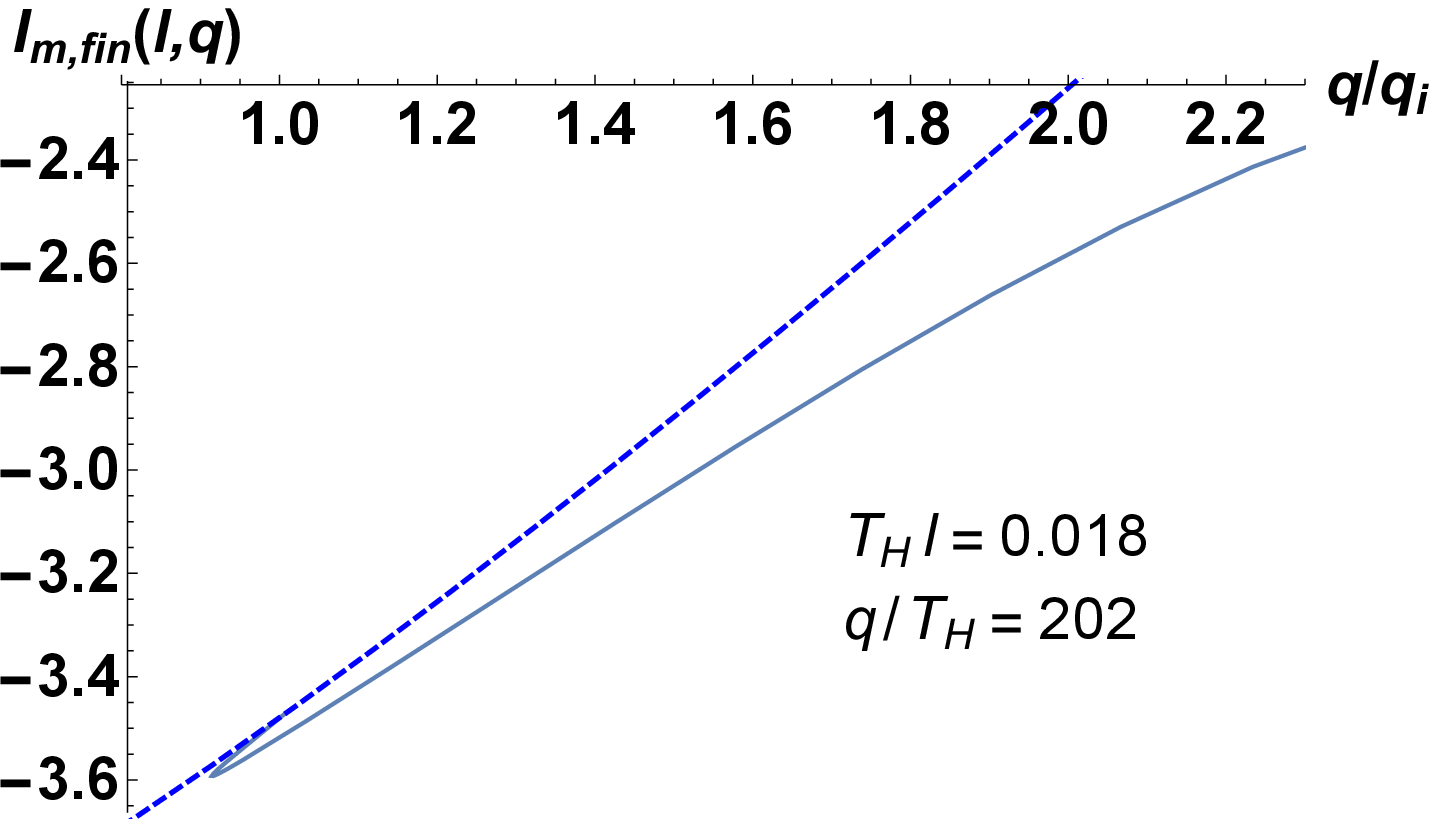} 
          \hspace{1.6cm}
  \includegraphics[width=2.5in]{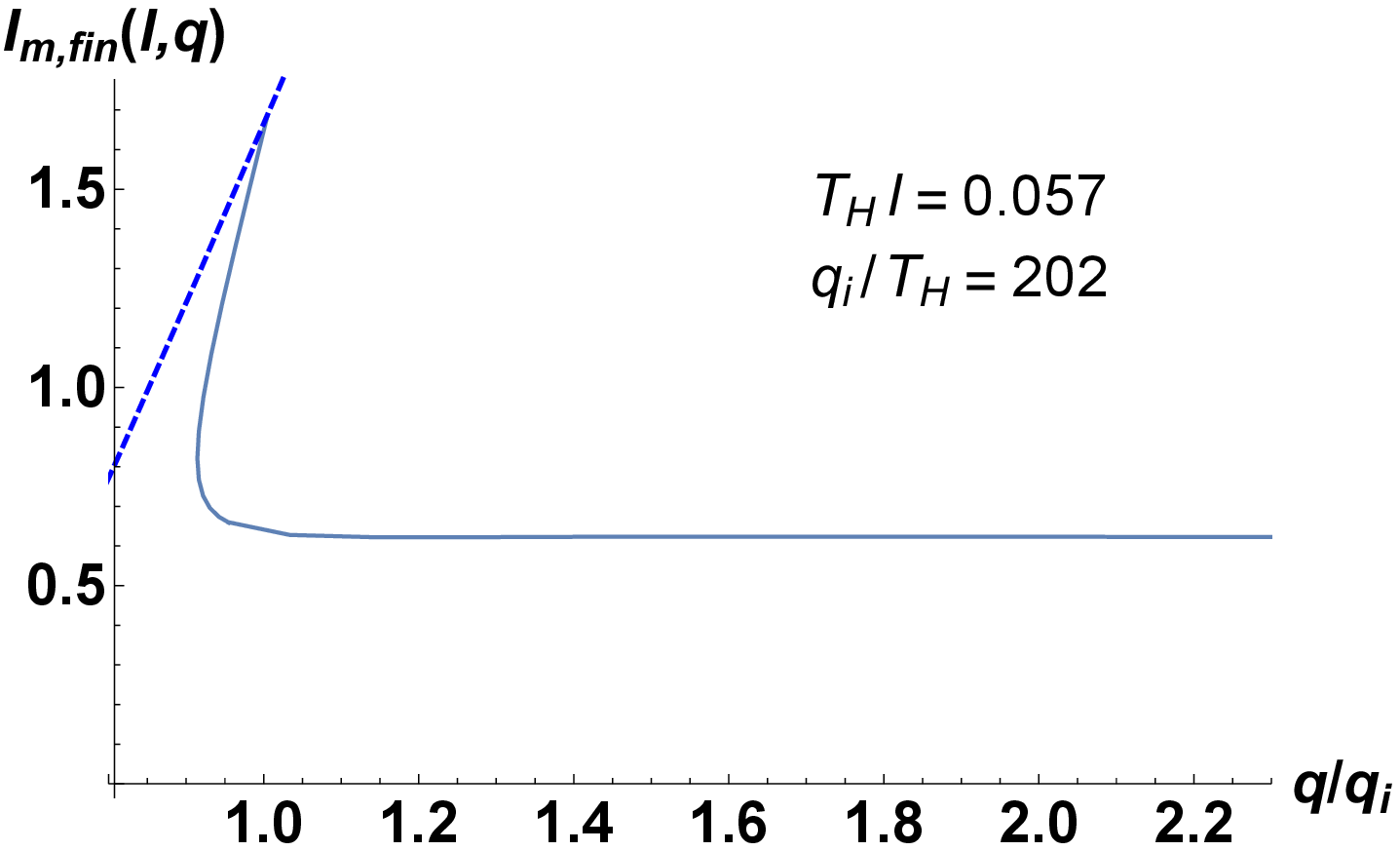} 
           \caption{We plotted the finite part of the holographic entanglement entropy $\kappa^2 S^{EE}/(2\pi)\equiv I_m$ as a function of $q/q_i$, when fixed $\tilde{\kappa}^2=0.5$.  The finite part in the condensed phase is always smaller than the one in the normal phase. The holographic entanglement entropy behaves non-monotonically. } 
    \label{fig:eeq}
    \end{center}
\end{figure}

\begin{figure}[htbp]
     \begin{center}
          \includegraphics[width=2.5in]{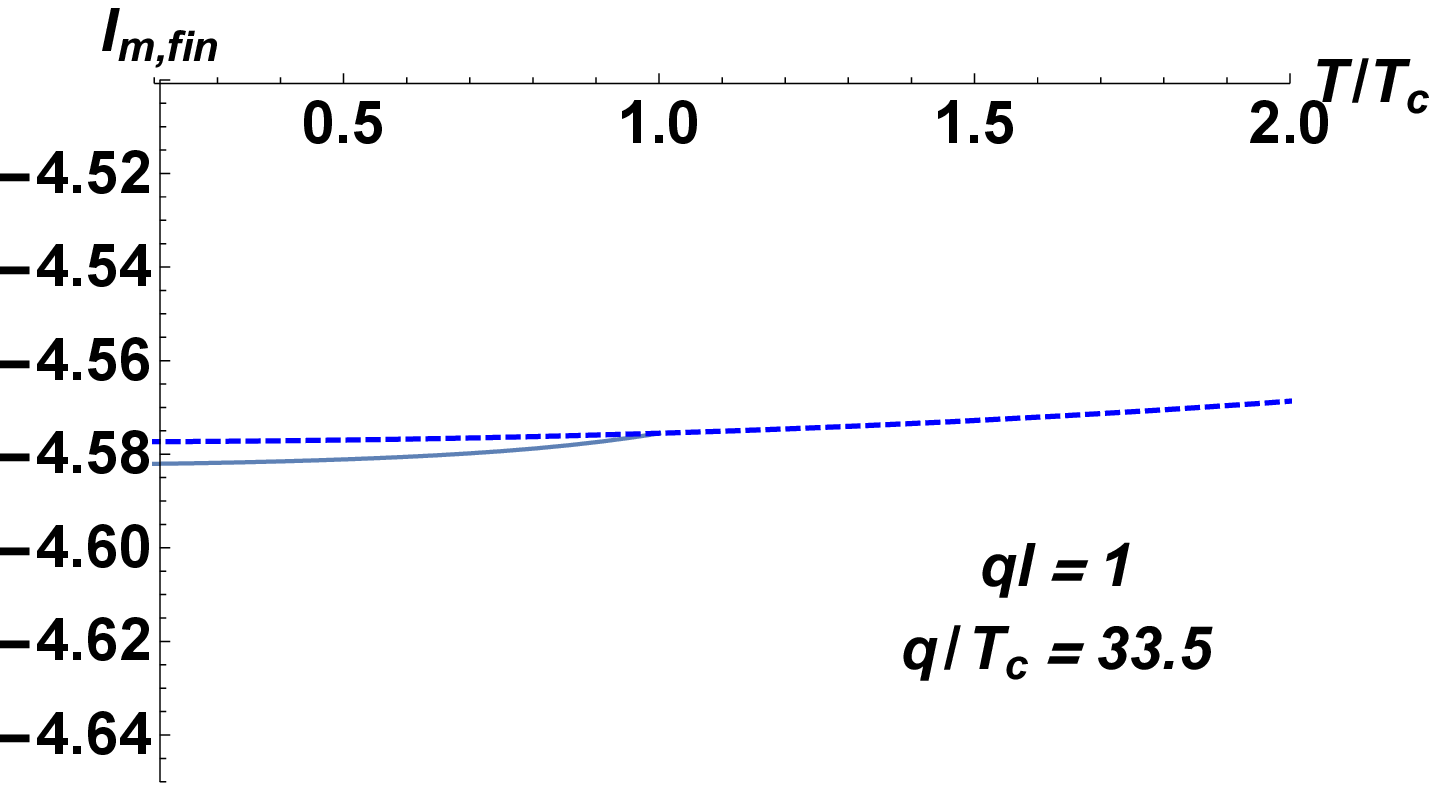} 
          \hspace{1.6cm}
  \includegraphics[width=2.5in]{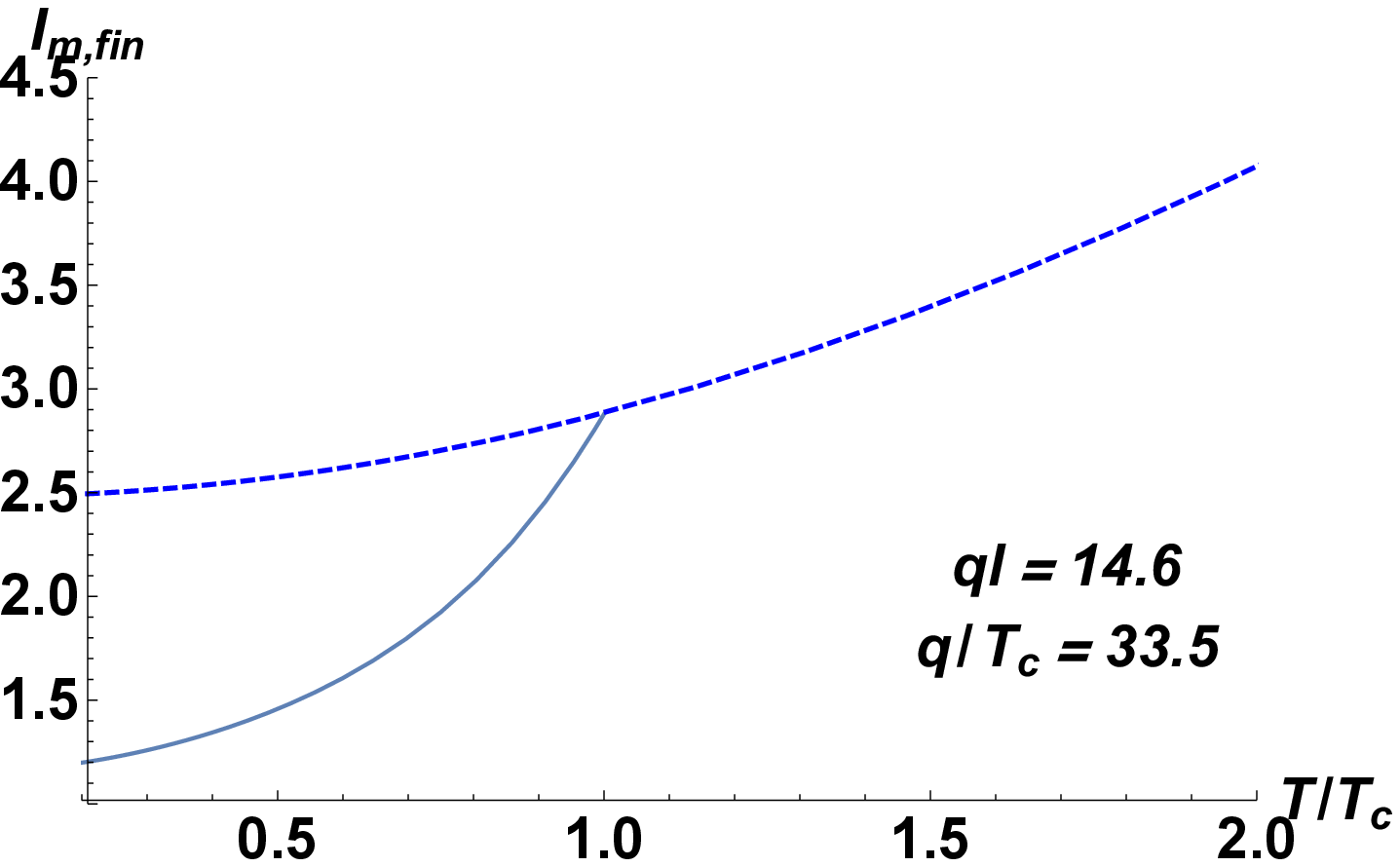} 
           \caption{We plotted the finite part of the holographic entanglement entropy $\kappa^2 S^{EE}/(2\pi)\equiv I_m$ as a function of $T/T_i$, for fixed $\tilde{\kappa}^2=0.1$ and the fixed size of the interval. By decreasing $T/T_c$, the superconductor phase appears below critical temperature. The finite part  in the condensed phase is always smaller than the one in the normal phase. } 
    \label{fig:eet}
    \end{center}
         \begin{center}
          \includegraphics[width=2.5in]{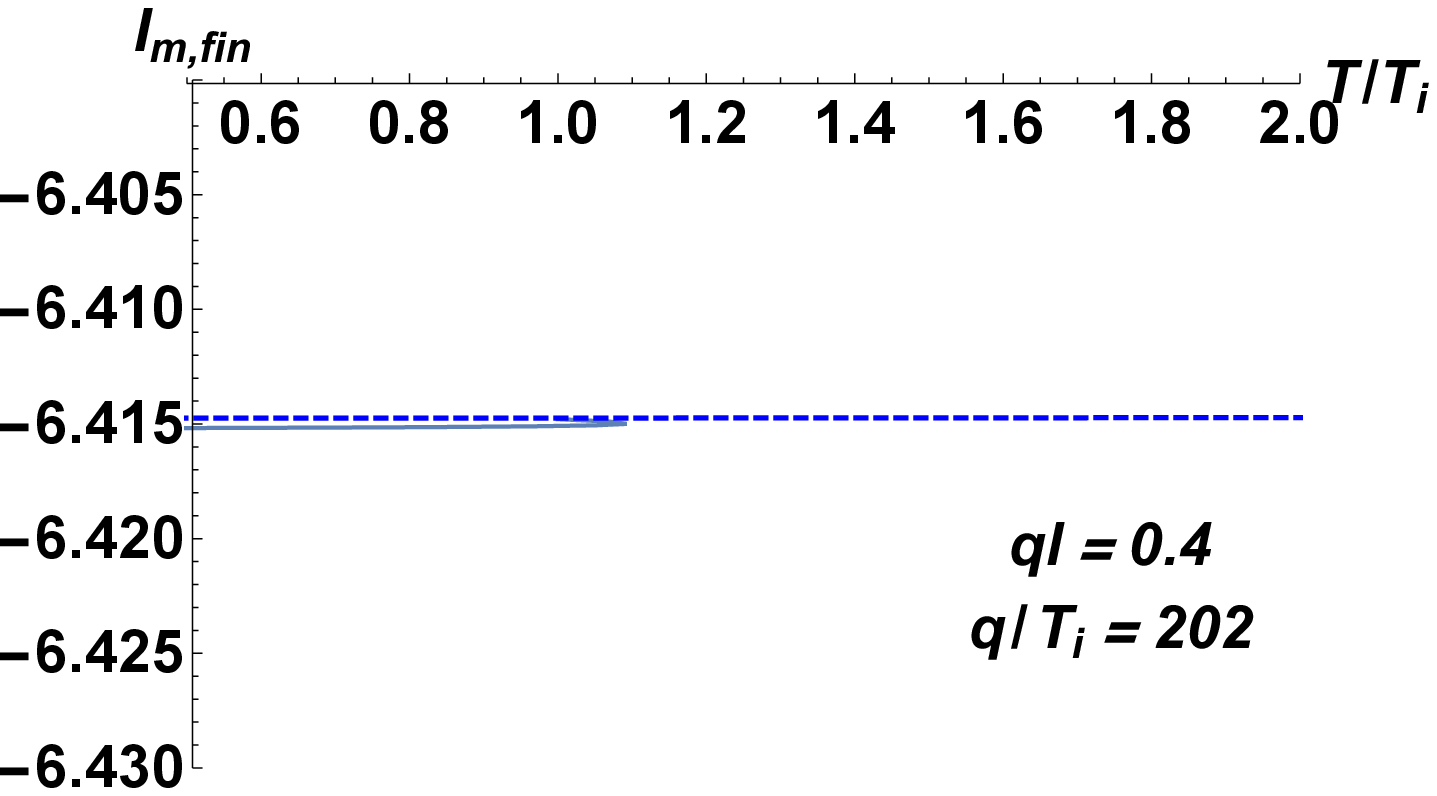} 
          \hspace{1.6cm}
  \includegraphics[width=2.5in]{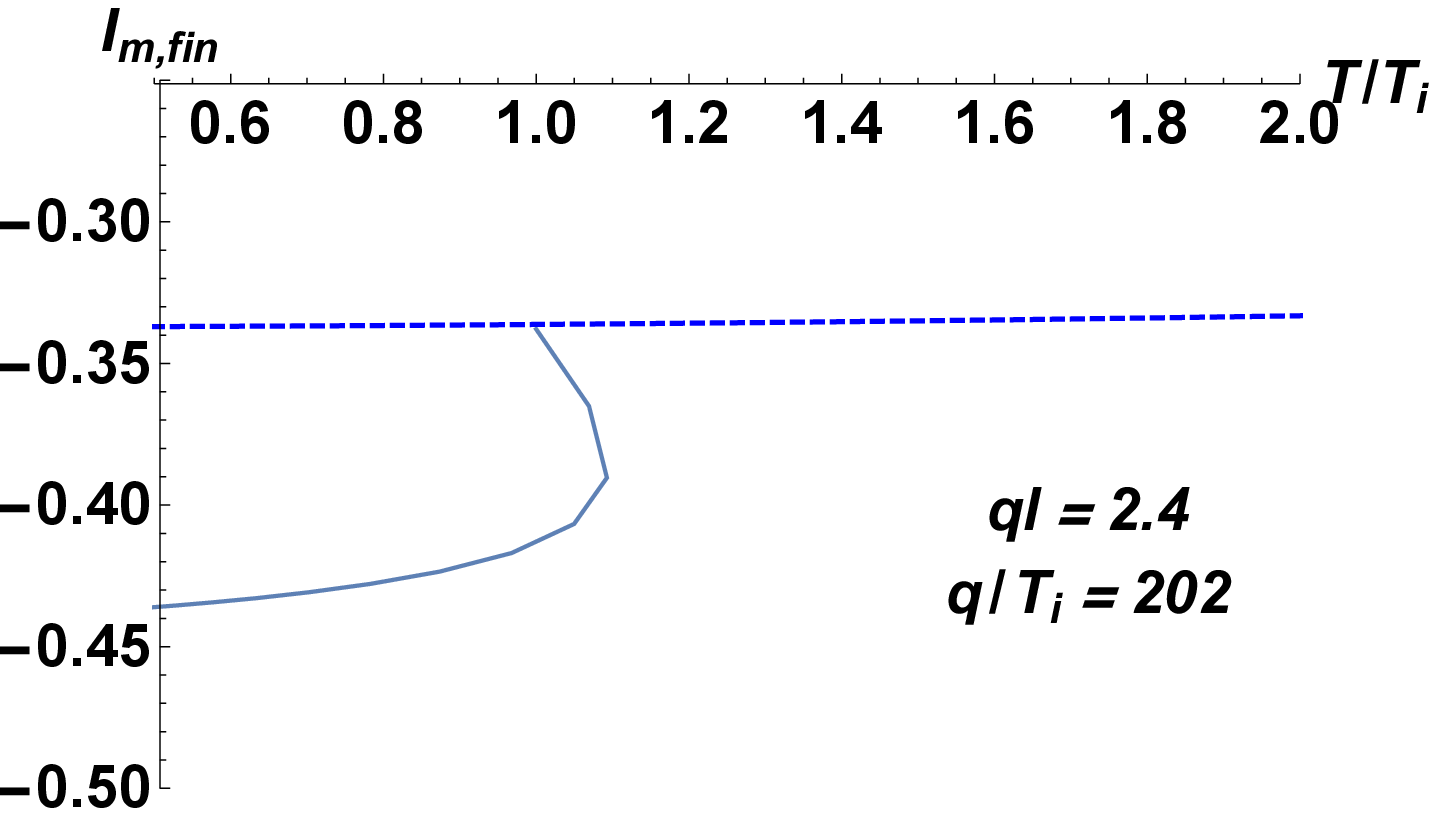} 
           \caption{The same quantity for fixed $\tilde{\kappa}^2=0.5$ and the fixed size of the interval. The finite part  in the condensed phase is always smaller than the one in the normal phase.  The finite part in the condensed phase turns out to be multi-valued around the critical point $T_c$.} 
    \label{fig:eet2}
    \end{center}
\end{figure}

In this section, we compute the finite part of the holographic entanglement entropy. 
Due to the subtraction of the divergent part  from $S^{EE}$, we define $I_{m,fin}\equiv \frac{\kappa^2}{2\pi}S^{EE}-2\log (\epsilon)$. When $\tilde{\kappa}^2=0.5$, the finite part of the holographic entanglement entropy is plotted as a function of $q/q_i$ in Fig. \ref{fig:eeq}. It has a cusp at the intersecting point between two curves of normal and condensed phases at the critical point. The finite part $I_{m,fin}$ behaves non-monotonically. Compared with  $\tilde{\kappa}^2<0.31$, the holographic entanglement entropy turns out to be multivalued at a region of small $q<q_i$. For large $q>q_i$, by contrast, the behavior of the holographic entanglement entropy is qualitatively similar to the one of $\tilde{\kappa}^2<0.31$. By increasing $l$, the increasing behavior of $I_{m,fin}(l,q)$ with increase of $q$ is varied into the decreasing behavior at large $q$. This phase transition occurs because $S^{EE}$ probes the formation of the condensate at large interval $l$. The decreased DOF due to the formation of the condensate overcomes increasing entanglement of charged states.

In figures \ref{fig:eet} and \ref{fig:eet2}, the finite part of the holographic entanglement entropy is plotted as a function of $T/T_i$. There is a cusp between two curves of normal and condensed phases. At low temperature, the finite part of the condensed phase is always lower than the one of the normal phase. When $\tilde{\kappa}^2=0.5>0.31$, the finite part turns out to be multi-valued around the critical point $T_c$.

Unlike the entanglement entropy as a function of $q$, we do not find any critical sizes where the finite part of the entanglement entropy has the phase transition. The finite part $I_{m,fin}$ decreases with decrease of $T/T_i$ at enough low temperature, while the finite part increases with increase of $q/q_i$ at high charge density for the small size. This implies that  by decreasing temperature, the amount of the quantum entanglement decreases. Simultaneously, the condensate is formed and degrees of freedom decreases~\cite{Das:2017gjy}.

\section{Holographic subregion complexity in asymptotically $AdS$ backgrounds}
In this section, we analytically compute the holographic subregion complexity in asymptotically $AdS$ backgrounds for comparison. The subregion complexity has a divergent part like $1/\epsilon$ and a finite part. 
\subsection{The subregion complexity of the pure $AdS$}
We consider Ryu-Takayanagi surface in the pure $AdS_3$ background (see section \ref{sec3} for Ryu-Takayanagi surface).  
Solving the EOM in terms of the embedding scalar $x(z)$ in the $AdS_3$ in the unit radius, it is evaluated as
\ba\label{C1}
x(z)=\int^{z_*}_{z}\dfrac{dz}{\sqrt{\dfrac{z_*^2}{z^2}-1}}=\sqrt{z_*^2-z^2},
\ea
where $z_*$ is the position of the turning point. The boundary condition $x(\epsilon)=l/2$ is imposed and then $l=2 z_*$.

Substituting \eqref{C1} into the holographic subregion complexity eq. \eqref{COM3}, we can analytically integrate it to give~\footnote{Following~\cite{Abt:2017pmf}, another definition of the subregion complexity turns out to be
\ba\label{CR2}
C_2=-\dfrac{1}{2}\int \sqrt{-g}R d^2x.
\ea
One can apply the Gauss-Bonnet theorem for the above formula. In asymptotically $AdS_3$ geometries, the divergent term is proportional to $l/\epsilon$. The finite term consists of the Euler number and the contribution coming from the extrinsic geodesic. The formula \eqref{CR2} gives the same result as eq. \eqref{COM3} when the Ricci scalar $R$ is a constant.
}
\ba\label{COMA6}
 C=\dfrac{c}{12\pi}\Big(\dfrac{l}{\epsilon}-{\pi}\Big),
\ea
where $c=12\pi/\kappa^2$ is the central charge. The subregion complexity is divergent like $1/\epsilon$.

By considering a rectangular shape of the surface $\gamma_A$, the constant term of eq. \eqref{COMA6} vanishes as follows:
\ba
C=\dfrac{c}{12\pi}\int^{l/2}_{-l/2} dx\int^{\epsilon}_{\infty} dz\dfrac{1}{z^2}=\dfrac{cl}{12\pi\epsilon}.
\ea

\subsection{The subregion complexity of the BTZ black hole}
We consider Ryu-Takayanagi surface in a BTZ black hole in the unit $AdS$ radius ($f(z)=1-z^2/z_h^2$ and $h(z)=1$ in eq. \eqref{MET1}).  Solving the EOM in terms of the embedding scalar $x(z)$, it can analytically be integrated as follows:
\ba\label{C4}
x(z)=\int^{z_*}_z\dfrac{dz}{\sqrt{f(z)\Big(\dfrac{z_*^2}{z^2}-1\Big)}}=z_h\coth^{-1}\Big(\sqrt{\dfrac{z_h^2-z^2}{z_*^2-z^2}}\Big),
\ea
where $z_*$ is the position of the turning point.
The boundary condition $x(\epsilon)=l/2$ shows $l=2 z_h \tanh^{-1}(z_*/z_h)$.

Substituting eq. \eqref{C4} into the subregion complexity eq. \eqref{COM3}, the complexity can be integrated out as follows:
\ba
{C}=\dfrac{c}{12\pi}\Big(\dfrac{l}{\epsilon}-{\pi}\Big).
\ea

The subregion complexity of the whole spatial circle is easily evaluated. It does not have a finite term as follows: 
\ba
C=\dfrac{c l}{12\pi \epsilon}.
\ea
At the large size $l$, Ryu-Takayanagi surface wraps the black hole horizon. Accordingly, the subregion complexity at a large interval eq. \eqref{COM39} has a finite term of the opposite sign as follows:
\ba
C=C_{entire}-C(2\pi -l)=\dfrac{c}{12\pi}\Big(\dfrac{l}{\epsilon}+\pi\Big).
\ea
Thus, the subregion complexity has its maximum similar to the holographic superconductor in main section when it wraps the almost whole spatial circle.

\end{document}